\documentclass[usenatbib]{mnras} 
\usepackage[utf8]{inputenc}
\usepackage[T1]{fontenc}
\usepackage[english]{babel}
\usepackage{amsmath}
\usepackage{amssymb}
\usepackage{graphicx}
\usepackage{lscape}
\usepackage[switch]{lineno}

\newcommand{\pd}[1]{\, \partial #1 \,}
\newcommand{\td}[1]{\, {\rm d} #1 \,}
\newcommand{\intl}{\int\limits}
\newcommand{\suml}{\sum\limits}
\newcommand{\SF}[3]{\; H\left[ #1;\, #2,\, #3 \right]}  

\newcommand{\g}{\ensuremath{\gamma}}
\newcommand{\zred}{z_{\rm red}}

\newcommand{\p}{^{\prime}}
\newcommand{\obs}{^{\rm obs}}
\newcommand{\est}[3]{\left( \frac{#1}{#2} \right)^{#3}}
\newcommand{\E}[1]{\times 10^{#1}}
\title[ExHaLe-jet]{\textit{ExHaLe-jet}: An extended hadro-leptonic jet model for blazars. I. Code description and initial results}

\author[M. Zacharias et al.]{
M. Zacharias$^{1,2}$\thanks{michael.zacharias@obspm.fr, mzacharias.phys@gmail.com}, A. Reimer$^3$, C. Boisson$^1$, A. Zech$^1$
\\
$^1$Laboratoire Univers et Théories, Observatoire de Paris, Université PSL, CNRS, Université de Paris, 92190 Meudon, France \\
$^2$Centre for Space Science, North-West University, Potchefstroom, 2520, South Africa \\
$^3$Institut f\"ur Astro- und Teilchenphysik, Leopold-Franzens-Universit\"at Innsbruck, A-6020 Innsbruck, Austria
}

\date{Accepted 2022 March 14. Received 2022 March 14; in original form 2021 September 3}

\pubyear{2021}

\begin{document}
\label{firstpage}
\pagerange{\pageref{firstpage}--26}
\maketitle

\begin{abstract}
The processes operating in blazar jets are still an open question. Modeling the radiation emanating from an extended part of the jet allows one to capture these processes on all scales.
Kinetic codes solving the Fokker-Planck equation along the jet flow are well suited to this task, as they can efficiently derive the radiation and particle spectra without the need for computationally demanding plasma-physical simulations. Here, we present a new extended hadro-leptonic jet code -- \textit{ExHaLe-jet} -- which considers simultaneously the processes of relativistic protons and electrons. Within a pre-set geometry and bulk flow, the particle evolution is derived self-consistently. Highly relativistic secondary electrons (and positrons) are created through \g-\g\ pair production, Bethe-Heitler pair production, and pion/muon decay. These secondaries are entrained in the jet flow decreasing the ratio of protons to electrons with distance from the jet base. For particle-photon interactions, we consider all internal and many external photon fields, such as the accretion disk, broad-line region, and the dusty torus. The external fields turn out to be the most important source for particle-photon interactions governing the resulting photon and neutrino spectra. In this paper, we present the code and an initial parameter study, while in follow-up works we present extensions of the code and more specific applications.
\end{abstract}




%
%
\section{Introduction}
The emission of blazars is typically modelled with the so-called one-zone model, where the emission of the relativistic jet is approximated as emanating from a single, small emission zone somewhere located in the jet. This approximation is justified by the significant variability observed on all time scales from years down to minutes, implying a size-limited emission region. It is, indeed, remarkable that the observed luminosity of blazars can vary by orders of magnitude, as for example in the sources PKS~2155-304 \citep{hess07pks}, 3C~454.3 \citep{vercellone+11}, 3C~279 \citep{fermi16,hess19}, PKS~1510-089 \citep{hess+21}, and CTA~102 \citep{zacharias+17}. In these examples, the one-zone model is clearly justified.

The location of the emission region is also debated for flaring events. The detection of flat spectrum radio quasars (FSRQs) at very-high-energy \g\ rays ($E>100\,$GeV) demands that the emission region be located outside of the broad-line region (BLR), even though inverse-Compton (IC) emission of the BLR has been the standard emission scenario for a long time \citep[e.g.,][]{hess19}. Additionally, the association of certain \g-ray flares with the ejection/motion of radio knots in the jets places the emission region of these flares from a few parsecs \citep{magic17} up to several tens of parsecs \citep{hess+21} away from the black hole.

In quiescent states, the one-zone approximation -- even though widely used -- may not be justified at all, as the lack of varaibility does not allow one to derive a limit on the size of the emission region. Furthermore, radio VLBI, optical, and X-ray observations of extended jet structures show that jets contain relativistic particles emitting synchrotron emission on all scales up to the termination point of the jet \citep[for a review on X-ray jets see, e.g.,][]{hk06}. Most notably, the detection of extended very-high-energy \g-ray emission along the jet of Centaurus~A \citep{hess20} indicates the presence of highly energetic particles on vast jet scales. Another interesting example is the blazar AP~Librae, where the \g\ rays cannot be successfully reproduced within a leptonic one-zone model requiring the need of an extended jet component to explain the \g-ray spectrum \citep{hervet+15,zw16,roychowdhury+22}. In turn, the one-zone model is a bad approximation for these resolved structures. 

This demands radiation models beyond the one-zone model. While MHD, RMHD and GRMHD codes have improved (and continue to do so) to model jets on vast scales \citep[e.g.,][]{chatterjee+20,dong+20,fichet+21}, the efficient calculation of all kinds of radiation processes \citep[for a recent review, see][]{cerruti20} is best done with kinetic models. In such models, the kinetic equation governing the particle distribution under influences of acceleration, cooling and other losses, is solved along the jet flow by cutting the jet into numerous slices and imposing a fixed jet geometry and bulk-flow evolution. While extended lepto-hadronic models exist \citep[e.g.,][]{vila+12,pepe+15,kantzas+21}, these are applied to X-ray binaries, such as Cygnus X-1, with specific characteristics and particularly good data sets. In the case of AGN, only leptonic multi-zone models have been considered \citep[e.g.,][]{pc13,malzac14,lucchini+19} implying that they only consider processes involving electrons and positrons. Notably, the radiation processes are synchrotron radiation and IC emission scattering ambient photon fields, such as the present synchrotron photons (synchrotron-self Compton, SSC), as well as photons from the accretion disk (AD), the BLR, the dusty torus (DT), the host galaxy, and the cosmic microwave background (CMB). Along with adiabatic cooling and assumptions on the acceleration process, these models can reproduce well the multiwavelength spectra of blazars, and other jetted systems such as black-hole X-ray binaries \citep[e.g.,][]{zdziarski+14}.

The possible association of neutrinos with blazars \citep{icecube+18,hovatta+21} has rekindled the interest in lepto-hadronic models, where also relativistic protons are permitted within the jet. While relativistic protons can emit synchrotron emission in high magnetic fields, they also interact in multiple ways with the ambient photon fields, most notably through Bethe-Heitler pair production and pion production. Charged pions decay into muons, which decay further into electrons and positrons. Neutral pions decay directly into photons exhibiting energies well in excess of hundreds of TeV. These photons can also interact with the ambient low-energy photon fields to produce pairs through \g-\g\ annihilation. As the pairs produced in all these processes are extremely relativistic, they in turn produce highly-energetic synchrotron and inverse-Compton emission initiating the so-called pair cascade, which is an avalanche of pairs. Meanwhile, the charged pions and muons -- while short-lived -- can also produce synchrotron emission.

However, applications of the lepto-hadronic one-zone model to the blazar TXS~0506-056 indicate that they cannot well reproduce the multiwavelength spectrum and the neutrino detection at the same time \citep[e.g.,][]{gao+19,cerruti+19,czbea21,reimer+19}. 
This is further evidence that modeling of blazars should go beyond the one-zone models, and shows that it is important to develop a radiation model that considers the extension of the jet, as well as relativistic protons. As mentioned above, the presence of energetic protons substantially increases the prospects for a pair cascade. As pairs are stable particles -- the jet medium is not thick enough for pair annihilation to be important -- they are carried along the jet increasing the leptonic content of the jet compared to the protons. This can help to explain the observed ratio of $\sim 20$ for the number of pairs to protons in the radio lobes \citep{sikora+20}. Naturally, this will depend on the ambient photon fields, and the external fields might be critical \citep{ghisellini+92,celottifabian93,sikoramadejski00,celottighiesellini08,ghisellinitavecchio10}.

In this paper, we present our newly developed extended hadro-leptonic jet code -- \textit{ExHaLe-jet} -- and provide a parameter scan in order to demonstrate its capabilities. The code description is separated into two sections. In section~\ref{sec:large}, we discuss the assumed bulk flow and geometrical structure. We also present the spatial evolution of the magnetic field, and our assumptions on the external photon fields. Section~\ref{sec:slices} describes the calculations performed in each slice, namely the ingredients and solutions of the Fokker-Planck equation, as well as the radiation and neutrino spectra. We then proceed with a first set of parameters to describe in detail the results produced by the code, along with a brief parameter study (section~\ref{sec:results}). Lastly, we summaries and provide an outlook in section~\ref{sec:discussion}.

As jets are supported and fed by the AD--black-hole system, the accretion and Eddington luminosities provide important markers on the jet power. Simulations of magnetically arrested disks (MADs) indicate that the MAD state can support a jet exceeding the accretion power \citep[e.g.,][]{tchekhovskoy+11} through the Blandford-Znajek process \citep{blandfordznajek77}, and might even support super-Eddington accretion rates for some time. However, these time scales are short compared to the jet's life-time. In most lepto-hadronic applications to blazars, the jet power (vastly) exceeds the Eddington luminosity \citep{zb15} requiring a careful consideration of the power demand in an extended lepto-hadronic model.
In order to do so, we describe the particle injection power as a fraction of the Eddington luminosity. This ensures a limited power budget, as we will show in section~\ref{sec:results}. Additional constraints are put on the geometry and the bulk flow. The current paradigm based on numerous VLBI maps, states that jets exhibit initially a parabolic geometry \citep{pushkarev+17} in which the bulk flow accelerates. At larger distances, the jet geometry is conical \citep[e.g.,][for a recent example]{casadio+21}, where the bulk flow is stationary. 

Throughout the paper, quantities in the observer's frame are marked with a superscript ``obs'', while quantities with a hat are in the frame of the host galaxy. Unmarked quantities are in the comoving frame of the jet/slice or are invariant. Positrons and electrons are collectively referred to as electrons in the remainder of this manuscript.

%
%
\section{Large-scale structure} \label{sec:large}
%
In this section, we define global parameters and settings as a function of distance from the base of the jet, such as the geometry, the bulk flow evolution of the jet, the evolution of the primary particle injection, the magnetic field, and the external photon fields. These quantities are then used to derive the particle distribution in each slice of the jet (Sec.~\ref{sec:slices}).

%
%
\subsection{Geometry} \label{sec:geometry}
We ignore general relativistic effects, and place the jet at the innermost stable circular orbit, $z_0=6R_g=6 GM_{0}/c^2$, of a Schwarzschild black hole with mass $M_0$. $G$ is the gravitational constant, and $c$ is the speed-of-light. The termination of our jet is set at $z_{\rm term}$, which is a free parameter. Between the jet base and termination, we construct a logarithmic grid $z_i$ implying the same number of slices per decade of distance. The $z$ coordinate, against which all parameters and equations are defined, is the arithmetic mean of subsequent grid points, $z = (z_i+z_{i+1})/2$. In turn, the length of a slice is given by $\Delta_{z}(z) = z_{i+1}-z_i$.

While any geometries can easily be implemented in our code, we here split the jet into a parabolic acceleration and a conical coasting section \citep[c.f.,][]{boccardi+21,park+21}. Following \cite{lucchini+19}, the bulk flow is accelerated by dissipating the (initially high) magnetic field. The evolution of the bulk Lorentz factor $\Gamma_b$ is then given as

\begin{align}
    \Gamma_b (z) = \begin{cases}
        \Gamma_{b,0} + (\Gamma_{b,{\rm max}}-\Gamma_{b,0}) \frac{\sqrt{z}-\sqrt{z_0}}{\sqrt{z_{\rm acc}}-\sqrt{z_0}} & z\leq z_{\rm acc} \\
        \Gamma_{b,{\rm max}} & z>z_{\rm acc}
    \end{cases}
    \label{eq:BLF}.
\end{align}
Here, $z_{\rm acc}$ is the length of the bulk acceleration region, $\Gamma_{b,{\rm max}}$ is the maximum bulk Lorentz factor, and $\Gamma_{b,0}=1.09$ its initial value. These are free parameters. The Doppler factor is given as $\delta(z) = [\Gamma_b(z) (1-\beta_b(z)\cos{\theta_{\rm obs}})]^{-1}$, with $\beta_b(z)=\sqrt{1-\Gamma_b(z)^{-2}}$, and the free parameter $\theta_{\rm obs}$ being the observation angle between the jet and the line-of-sight.

Apparently, a strong connection exists between the jet's opening angle and the bulk flow \citep[e.g.,][]{pushkarev+09}. Therefore, we set the radius of the jet as

\begin{align}
    R(z) = \eta_R z_0 + (z-z_0) \tan{(\eta_{o}/\Gamma_b(z))}
    \label{eq:jetradius},
\end{align}
with the free parameters $\eta_R$ defining the minimum jet radius as a multiple of the jet base, and the multiple $\eta_{o}$ of the opening angle. For the latter, \cite{pushkarev+17} found a median value of $\eta_o = 0.26$.

%
%
\subsection{Primary particle injection} \label{sec:primaryparticleinjection}
At the jet base, we inject a plasma of protons and electrons (including positrons, unless the distinction is necessary) with injection luminosity $L_{\rm inj}$:

\begin{align}
    L_{\rm inj} &= q(z_0)\pi R(z_0)^2 \Delta_{z}(z_0) \nonumber \\
    &\quad\times \left[ m_pc^2 \mathcal{I}_{p1}+\frac{m_ec^2}{\mathcal{I}_{e0}\kappa_{pe}(z_0)} \mathcal{I}_{e1}\mathcal{I}_{p0} \right]
    \label{eq:injlum00}.
\end{align}
Here, $q(z_0)$ is the injection rate at the base, $m_i$ are the particle masses for species $i$ (protons and electrons, in this case), and $\kappa_{pe}(z_0)$ is the proton to electron density ratio at the jet base. Protons and electrons are injected with a power-law distribution (index $p_i$) with respect to the particle Lorentz factor $\gamma$ between a minimum and a maximum value $\gamma_{i,1}$ and $\gamma_{i,2}$, respectively. The integrals over the injection distributions are:

\begin{align}
    \mathcal{I}_{ik} = \intl_{\gamma_{i,1}}^{\gamma_{i,2}} \gamma^{k-p_i} \td{\gamma}
    \label{eq:injectintegrals}.
\end{align}
A detailed derivation of Eq.~(\ref{eq:injlum00}) is given in App.~\ref{app:partmag}. 

The jet power is provided by a fraction of the accretion power. The corresponding accretion disk radiates with a luminosity that is a fraction $l_{\rm edd}$ of the Eddintion power, $\hat{L}_{\rm AD}=l_{\rm edd}L_{\rm edd}$ with $L_{\rm edd} = 4\pi GM_0m_pc/\sigma_T$, and $\sigma_T$ being the Thomson cross section. 
Relating $L_{\rm inj}$ with the accretion dynamics, we can write

\begin{align}
    L_{\rm inj} = \frac{f_{\rm inj}L_{\rm edd}}{2\Gamma_{b,0}^2}
    \label{eq:injlum01},
\end{align}
where $f_{\rm inj}$ is a free parameter that determines the power that is injected into \textit{two} jets in the form of particles. In this work, typically, the power injected into the magnetic field is larger than that injected as particles. Hence, $f_{\rm inj}\ll 1$ in cases where the Eddington luminosity limits the jet power (cf. App.~\ref{app:partmag}). 
%
It is convenient to define $L_{\rm inj}$ as a function of $L_{\rm edd}$, as we treat the accretion luminosity $L_{\rm AD}$ as a free parameter. A comparison of the jet power to $L_{\rm AD}$ will be provided in Sec.~\ref{sec:results}.
Combining Eqs.~(\ref{eq:injlum00}) and (\ref{eq:injlum01}) provides the initial injection rate $q(z_0)$. 

Under the assumption of conserved particle flux, the injection rate of each slice obeys the continuity equation \citep[c.f.,][]{pc13}:
\begin{align}
     \frac{\td{}}{\td{z}} \left[ \Gamma_b(z)\beta_b(z)q(z)R(z)^2\Delta_{z}(z) \right] = 0
     \label{eq:continuity}
\end{align}
Hence, at distance $z$ from the jet base $z_0$ the normalization factor becomes

\begin{align}
     q(z) = q(z_0) \frac{\Gamma_b(z_0)\beta_b(z_0)\Delta_{z}(z_0)}{\Gamma_b(z)\beta_b(z)\Delta_{z}(z)} \left( \frac{R(z_0)}{R(z)} \right)^2
     \label{eq:p-inject1}.
\end{align}
As further discussed in Sec.~\ref{sec:slices}, the conservation of particle flux is approximately true for protons, as we neglect at this point the conversion of protons into neutrons and back. In case of significant secondary pair injection, the electron flux is not conserved globally. Nonetheless, applying Eq.~(\ref{eq:continuity}) from one slice to the next provides the injection rate for electrons in subsequent slices.

%
%
\subsection{Magnetic field evolution and injection constraints} \label{sec:magfield}
The magnetic field $B(z)$ is evolved following the relativistic Bernoulli equation \citep{koenigl80,Zdziarski+15}:

\begin{align}
    \Gamma_b(z)\left[ 1+\frac{\eta_{\rm ad} u(z)+B(z)^2/4\pi}{\rho(z) c^2} \right] = \mbox{const}
    \label{eq:Bernoulli},
\end{align}
where we set the adiabatic index to its relativistic value $\eta_{\rm ad}=4/3$, $u(z)=u_p(z)+u_e(z)$ is the sum of the proton and electron energy densities, while $\rho(z)=[m_pn_p(z)+m_en_e(z)]$ is the sum of the proton and electron rest mass densities. 
Equating the left-hand-side of Eq.~(\ref{eq:Bernoulli}) to the respective fraction taken at the base ($z=z_0$), we can solve for the magnetic field $B(z)$ depending on the initial magnetic field $B(z_0)$, which is a free parameter. 

An unperturbed flow -- that is, assuming negligible energy gains and losses, as well as no secondary injection -- can be calculated with the equations given in Sec.~\ref{sec:primaryparticleinjection}. This allows us to calculate the magnetic field along the jet first, and to impose this ``unperturbed'' magnetic field on the jet. Hence, we fix the geometry, the bulk flow evolution and the magnetic field, and then subsequently allow the particle distributions to evolve (including pair creation) as described in Sec.~\ref{sec:slices}.

With the help of the magnetization, that is the ratio of magnetic to particle enthalpy,

\begin{align}
    \sigma_{B}(z) = \frac{B(z)^2/4\pi}{\eta_{\rm ad} u(z)+\rho(z)c^2}
    \label{eq:magnetization},
\end{align}
we can rewrite Eq.~\ref{eq:Bernoulli} as

\begin{align}
    \Gamma_b(z)\left[ 1+\sigma_B(z) \right]\,\left[ 1+\frac{\eta_{\rm ad}u(z)}{\rho(z)c^2} \right] = \mbox{const}
    \label{eq:bernoulli2}.
\end{align}
Using the initial value at $z=z_0$ as the constant, we can solve the resulting equation for $\sigma_B(z)$ resulting in

\begin{align}
    \sigma_B(z) &= \left[ 1+\sigma_B(z_0) \right] \frac{\Gamma_{b,0}}{\Gamma_b(z)} \frac{1+\frac{\eta_{\rm ad}u(z_0)}{\rho(z_0)c^2}}{1+\frac{\eta_{\rm ad}u(z)}{\rho(z)c^2}} - 1 \nonumber \\
    &\approx \left[ 1+\sigma_B(z_0) \right] \frac{\Gamma_{b,0}}{\Gamma_b(z)} - 1
    \label{eq:bernoulli3},
\end{align}
where the approximation holds for the ``unperturbed'' flow. Demanding at the termination point of our jet, $z=z_{\rm term}$, $\sigma_B(z_{\rm term})>0$ immediately leads to the initial condition $\sigma_B(z_0)>(\Gamma_{b,{\rm max}}/\Gamma_{b,0})-1$. As demonstrated in App.~\ref{app:partmag}, this condition restricts the injection fraction $f_{\rm inj}$ to

\begin{align}
    f_{\rm inj} &< \frac{c\Gamma_{b,0}^2\eta_R^2z_0^2B(z_0)^2}{4L_{\rm edd}} \frac{2}{\left( \frac{\Gamma_{b,{\rm max}}}{\Gamma_{b,0}}-1 \right)\eta_{\rm esc}\zeta} \nonumber \\
    &= 1.0\E{-5} \est{\Gamma_{b,0}}{1.09}{2} \est{\eta_R z_0}{10^{15}\,\mbox{cm}}{2} \est{B(z_0)}{50\,\mbox{G}}{2} \nonumber \\
    &\quad\times \est{M_0}{10^{8}\,M_{\odot}}{-1} \est{\Gamma_{b,{\rm max}}}{30}{-1} \est{\eta_{\rm esc}}{10}{-1} \est{\zeta}{4/3}{-1}
    \label{eq:fedd},
\end{align}
which solely depends on input parameters. Here, $\eta_{\rm esc}>1$ is a multiple of the light crossing time scale (see below), and $\zeta>4/3$ is a function of the particle distributions [see Eq.~(\ref{eq:app_zeta})]. Equation~(\ref{eq:fedd}) implies that the jet can only support a certain maximum initial particle density in order to satisfy Eq.~(\ref{eq:Bernoulli}) for a given bulk Lorentz factor and magnetic field.

%
%
\subsection{External photon fields} \label{sec:extfield}
We employ four external photon fields: the AD, the BLR, the DT, and the CMB. For the AD, we use the standard thin-disk model of \cite{ss73} extending between the innermost stable orbit 
%
and a maximum radius defined by the point where the AD becomes unstable due to self-gravity \citep[e.g.,][his Eq.~(2)]{netzer15},

\begin{align}
    \hat{R}_{\rm AD,max} = 1680R_g \left( \frac{M_0}{10^9 M_{\odot}} \right)^{-2/9} \alpha^{2/9} l_{\rm edd}^{4/9} \left( \frac{\xi}{0.1} \right)^{-4/9}
    \label{eq:ADrmax}.
\end{align}
Here, 
$\alpha$ is the disk's viscosity, while $\xi$ is the mass-to-radiation conversion efficiency. We set $\alpha=\xi=0.1$. 

The temperature profile of the AD as a function of disk radius $\hat{r}_{\rm AD}$ is

\begin{align}
    \hat{T}_{\rm AD}(\hat{r}_{\rm AD}) = \left( \frac{3GM_0 \hat{L}_{\rm AD}}{8\pi \xi c^2 \sigma_T \hat{r}_{\rm AD}^3} \right)^{1/4}
    \label{eq:ADTemp}.
\end{align}
In terms of the normalized disk temperature $\hat{\Theta}_{\rm AD}=k_B\hat{T}_{\rm AD}/m_ec^2$, where $m_e$ is the electron rest mass and $k_B$ the Boltzmann constant, the observed spectral luminosity of the AD becomes

\begin{align}
    \nu\obs L\obs_{\nu\obs} &= \frac{l_{\rm edd}L_{\rm edd}\cos{\theta_{\rm obs}}}{2\xi (R_{\rm AD,min}/R_g)} \left( \frac{\hat{\Theta}_{\rm AD,min}}{\hat{\Theta}_{\rm AD,max}} \right)^{4/3} \nonumber \\
    &\times \frac{\left( \frac{h\nu\obs (1+z_{\rm red})}{m_ec^2 \hat{\Theta}_{\rm AD,min}} \right)^{4}}{1+\left( \frac{h\nu\obs (1+z_{\rm red})}{m_ec^2 \hat{\Theta}_{\rm AD,min}} \right)^{8/3}} \exp{\left\{ -\frac{h\nu\obs (1+z_{\rm red})}{m_ec^2 \hat{\Theta}_{\rm AD,max}} \right\}}
    \label{eq:ADluminosity},
\end{align}
with the cosmological redshift $z_{\rm red}$, and employing $\hat{\Theta}_{\rm AD,min}\equiv \hat{\Theta}_{\rm AD}(\hat{r}_{\rm AD,min})$ and $\hat{\Theta}_{\rm AD,max}\equiv \hat{\Theta}_{\rm AD}(\hat{r}_{\rm AD,max})$. 

The BLR and the DT are approximated as a grey body radiation fields at temperature $\hat{T}_{\rm BLR}$ and $\hat{T}_{\rm DT}$, respectively, which are free parameters. We use the relations given in \cite{gt08} to obtain

\begin{align}
    \hat{R}_{\rm BLR} &= 10^{17} \left( \frac{l_{\rm edd}L_{\rm edd}}{10^{45}\,\mbox{erg/s}} \right)^{1/2}\,\mbox{cm} \label{eq:RBLR} \\
    \hat{R}_{\rm DT} &= 2.5\times 10^{18} \left( \frac{l_{\rm edd}L_{\rm edd}}{10^{45}\,\mbox{erg/s}} \right)^{1/2}\,\mbox{cm} \label{eq:RDT} 
\end{align}
for the BLR and the DT, respectively. The luminosities are generically set to $10\%$ of the accretion disk luminosity. However, in order to preserve the isotropy approximation in the galaxy frame for these photon fields, we assume the following dependence of the luminosities on distance $z$ \citep{hayashida+12}

\begin{align}
    \hat{L}_{\rm BLR} &= \frac{0.1l_{\rm edd}L_{\rm edd}}{(1+z/\hat{R}_{\rm BLR})^{3}} \label{eq:LBLRz} \\
    \hat{L}_{\rm DT} &= \frac{0.1l_{\rm edd}L_{\rm edd}}{(1+z/\hat{R}_{\rm DT})^{4}} \label{eq:LDTz},
\end{align}
respectively. The spectral luminosities in the observer's frame are

\begin{align}
    \nu\obs L\obs_{\nu\obs} &= \frac{0.1l_{\rm edd}L_{\rm edd}}{6} \left( \frac{h\nu\obs (1+z_{\rm red})}{k_B \hat{T}_{\rm BLR}} \right)^{4} \nonumber \\
    &\times \exp{\left\{ -\frac{h\nu\obs (1+z_{\rm red})}{k_B \hat{T}_{\rm BLR}} \right\}} \label{eq:BLRluminosity} \\
    \nu\obs L\obs_{\nu\obs} &= \frac{0.1l_{\rm edd}L_{\rm edd}}{6} \left( \frac{h\nu\obs (1+z_{\rm red})}{k_B \hat{T}_{\rm DT}} \right)^{4} \nonumber \\
    &\times \exp{\left\{ -\frac{h\nu\obs (1+z_{\rm red})}{k_B \hat{T}_{\rm DT}} \right\}} \label{eq:DTluminosity}, 
\end{align}
respectively. The numerical prefactor $1/6$ normalizes the energy spectra ensuring that the integral over $L\obs_{\nu\obs}$ provides the expected total luminosity, namely $0.1l_{\rm edd}L_{\rm edd}$.

The external photon fields have two impacts. Firstly, they serve as target photons for proton-photon and electron IC interactions, while secondly they act as absorbers of \g\ rays through pair production. Within the slices, these pairs add to the particle content as described below. Outside the jet, we only consider the absorption process resulting in a decrease of \g\ rays from a given slice, if it is located at $z<R_{\rm BLR,DT}$. For the external absorption, we use the code developed by \cite{be16}, where the BLR is represented by a quasar template spectrum normalized to the BLR luminosity, $\hat{L}_{\rm BLR}$, while for the DT a simple grey-body spectrum is used.



%
%
\section{Sliced calculations} \label{sec:slices}
Having imposed the geometry, the bulk flow, and the magnetic field on the jet, we can now proceed and calculate the ``perturbed'' particle distributions and the resulting photon and neutrino fluxes. 
In each slice at distance $z$ from the black hole, the particle distributions are calculated employing a Fokker-Planck equation. As the following equations are the same in every slice, we omit the explicit dependence on $z$ from the respective variables, unless the dependence is explicitly required. 

We solve the Fokker-Planck equation for four particle species, namely protons, pions, muons and electrons. The particle momentum is given by $p_i = \chi m_ic$, where $m_i$ is the particle mass, and $\chi=\gamma\beta$ with $\gamma = \sqrt{\chi^2+1}$ the particle's Lorentz factor and $\beta = v/c = \chi/\sqrt{\chi^2+1}$ its speed normalized to the speed of light. The Fokker-Planck equation then reads

\begin{align}
     \frac{\pd{n_i(\chi,t)}}{\pd{t}} = \frac{\pd{}}{\pd{\chi}} \left[ \frac{\chi^2}{(a+2)t_{\rm acc}} \frac{\pd{n_i(\chi,t)}}{\pd{\chi}} \right] \nonumber \\
	 - \frac{\pd{}}{\pd{\chi}} \left( \dot{\chi}_i n_i(\chi,t) \right) + Q_i(\chi)
	 - \frac{n_i(\chi,t)}{t_{\rm esc}} - \frac{n_i(\chi, t)}{\gamma t^{\ast}_{i,{\rm decay}}}. 
	 \label{eq:fpgen}
\end{align}
Here, $n_i$ is the particle density, $a$ is the ratio of shock to Alfv\`{e}n speed, $t_{\rm acc}$ the energy-independent acceleration time scale, $\dot{\chi}_i$ the momentum gain and loss rate, $Q_i$ the particle injection rate, $t_{\rm esc}$ the energy-independent particle escape time scale, and $t^{\ast}_{i,{\rm decay}}$ the decay time scale of the unstable particles in their frame of rest. The escape time scale in each slice is given by $t_{\rm esc} = \eta_{\rm esc}\Delta_{z}/c$, and $\eta_{\rm esc}>1$ a free parameter parameterizing the advective motion of particles in each slice. In App.~\ref{app:solver}, we describe the numerical scheme to solve Eq.~(\ref{eq:fpgen}).

%
%
\subsection{Particle injection terms}
We consider in each slice the injection of primary and secondary particles. We denote primaries as particles propagating from the upstream into the slice at hand, while secondaries are produced in the slice itself. Protons are primary particles, and we assume that the total number of protons in the jet is conserved. 
Hence, protons follow Eq.~(\ref{eq:p-inject1}).

Pions and muons decay rapidly, and we assume that they will not propagate through the jet but remain in the slice where they have been created. Hence, pions and muons are considered to be secondary particles only.

For electrons the situation is more complicated. We inject at the base of the jet a population of electrons along with the protons. To repeat, we denote with $\kappa_{pe}(z) = n_p(z)/n_e(z)$ the number density ratio of protons to electrons, which is a free parameter at the base of the jet, $z=z_0$. In each slice, pion production (followed by pion and muon decay), Bethe-Heitler pair production and \g-\g\ pair production create secondary electrons. As electrons are stable particles, they propagate downstream implying a decrease of $\kappa_{pe}(z)$.

Currently, we do not explicitly consider neutrons. The production of neutrons through proton-photon interactions and their subsequent evolution would not allow us to conserve the proton number, which is however necessary to use Eq.~(\ref{eq:p-inject1}) as is. Inclusion of the evolution of neutrons is planned for a future update of the code. 



%
\subsubsection{Primary injection}
In each slice, the primary proton and electron injection functions take the form of a power-law:

\begin{align}
     Q_i(\chi) = q_{i} \chi^{-p_i} \SF{\chi}{\chi_{i,1}}{\chi_{i,2}} 
     \label{eq:prim-inject},
\end{align}
where the spectral index $p_i$, and the lower and upper cut-offs, $\chi_{i,1}$ and $\chi_{i,2}$, respectively, are free parameters. Currently, we assume these to be the same in each slice. In future applications of the code, we plan to include a self-consistent evolution of these parameters along the jet. We note that the upper cut-off is reduced, if the Larmor radius exceeds the radius of a given slice; that is, we demand $\chi_{i,2}\leq (e/m_ic^2)BR$. 

With the help of Sec.~\ref{sec:primaryparticleinjection} and App.~\ref{app:partmag}, we can derive the normalization factors $q_{p}$ for protons and $q_{e}$ for electrons, respectively. For protons, the evolution along the jet is given by Eq.~(\ref{eq:p-inject1}), while electrons get an additional update from the evolution of $\kappa_{pe}$ ensuring that the created pairs are transported downstream.

\subsubsection{Secondary injection}
%
For the pion production, we use the template approach of \cite{huemmer+10} approximating the cross section by piece-wise step functions. The strict separation into the different interaction channels (ITs) -- such as $\Delta(1232)$-resonance, higher resonances, direct and multi-pion production -- is a simplification, but allows for the tabulation of the cross section $\sigma^{\rm IT}$ providing excellent agreement with the results of the SOPHIA Monte Carlo code \citep{muecke+00}. Hence, the pion injection rate becomes a sum over the different ITs:

\begin{align}
	Q_{\pi^i}(\chi) &= m_{\pi^i}c^2 \suml_{\rm IT} n_p\left( \frac{E_{\pi^i}}{\epsilon^{\rm IT}} \right) \frac{m_pc^2}{E_{\pi^i}} \nonumber \\
	&\quad\times \intl_{\epsilon_{thr}/2}^{\infty}\td{y} n_{\rm ph}\left( \frac{m_pc^2y\epsilon^{\rm IT}}{E_{\pi^i}} \right) M_{\pi^i}^{\rm IT} f^{\rm IT}(y)
	\label{eq:pioninject1}.
\end{align}
The injection is derived separately for the three pion types, namely $\pi^+$, $\pi^-$, and the neutral $\pi^0$. The lower limit of the integral marks the threshold beneath which the cross section is zero. The threshold is $\epsilon_{thr}=294$ corresponding to an energy of $150\,$MeV. The proton distribution $n_p$ is evaluated at the pion energy $E_{\pi^i}=\sqrt{\chi^2+1}m_{\pi^i}c^2$ divided by the mean energy fraction $\epsilon^{\rm IT}$ that is deposited into the daughter particles for a given interaction channel. The integration variable is $y=\sqrt{\chi_p^2+1}\epsilon$ relating the normalized proton energy with the normalized photon energy $\epsilon = h\nu/m_ec^2$. The photon distribution is described by $n_{\rm ph}(\epsilon)$.\footnote{We consider all photon fields -- internal and external -- as target photons in the particle-photon interactions. External photon fields are boosted into the comoving frame and then angle-averaged. While the latter is a simplification, it eases the computational effort with reasonable accuracy.} 
The functions $M_{\pi^i}^{\rm IT}$ and $f^{\rm IT}(y)$ represent the multiplicity of daughter particles and the simplified response function, respectively, of the interaction channel. The functions $\epsilon^{\rm IT}$, $M_{\pi^i}^{\rm IT}$, and $f^{\rm IT}(y)$ have been tabulated by \cite{huemmer+10}, which allows for a swift evaluation of Eq.~(\ref{eq:pioninject1}).

Given that neutral pions decay into photons within a proper time of $t^{\ast}_{\pi^0,{\rm decay}}=2.8\times 10^{-17}\,$s, their decay is basically instantaneous and we derive their electromagnetic emission directly from the injection spectrum (cf. Sec.~\ref{sec:radterm}). Charged pions decay in a proper time of $t^{\ast}_{\pi^i,{\rm decay}}=2.6\times 10^{-8}\,$s, which is long enough to potentially undergo changes in their energy distribution \citep[e.g.,][]{muecke+03}. Therefore, we solve Eq.~(\ref{eq:fpgen}) separately for the charged pion species and calculate their synchrotron emission. 

The charged pions decay into muons and neutrinos: 

\begin{align}
    \pi^+ &\rightarrow \mu^+ + \nu_{\mu} \label{eq:pionpdecay} \\
    \pi^- &\rightarrow \mu^- + \bar{\nu}_{\mu} \label{eq:pionmdeday}
\end{align}
providing the muon injection term 

\begin{align}
    Q_{\mu^i}(\chi) = \frac{n_{\pi^i}(\chi)}{\gamma t^{\ast}_{\pi^i,{\rm decay}}}
    \label{eq:muoninjection}.
\end{align}
We again solve Eq.~(\ref{eq:fpgen}) separately for the muons and calculate their synchrotron emission. Muons decay after a proper time of $t^{\ast}_{\mu^i,{\rm decay}}=2.2\times 10^{-6}\,$s into electrons or positrons and related neutrinos: 

\begin{align}
    \mu^+ &\rightarrow e^+ + \nu_e + \bar{\nu}_{\mu} \label{eq:muonpdecay} \\
    \mu^- &\rightarrow e^- + \bar{\nu}_e + \nu_{\mu} \label{eq:muonmdecay} 
\end{align}
Following \cite{rsch02}, we use

\begin{align}
    Q_{\rm +-}(\chi) &= \intl_1^{104}\td{\gamma_e} \frac{\gamma_e^2 (3-2\gamma_e/104)}{104^3\sqrt{\gamma_e^2-1}} \nonumber \\ 
    &\quad\times \intl_{\gamma\gamma_e(1-\beta\beta_e)}^{\gamma\gamma_e(1+\beta\beta_e)} \td{\gamma_{\mu}} \frac{n_{\mu^+}(\chi_{\mu})+n_{\mu^-}(\chi_{\mu})}{\gamma_{\mu}t^{\ast}_{\mu^i,{\rm decay}} \sqrt{\gamma_{\mu}^2-1}}
    \label{eq:muondecayinj},
\end{align}
where $\gamma_{e,{\rm max}}=104$ is derived from the kinematics of the process in the muon rest frame. We describe the calculation of the neutrino spectra in section~\ref{sec:neu}.

Electrons and positrons are also produced through Bethe-Heitler pair production. Following \cite{ka08}, the electron-positron injection rate for $\chi_p\gg 1$ can be written as

\begin{align}
	Q_{\rm BH}(\chi_e) &= 2c\intl_{1}^{\infty}\td{\gamma_p}\frac{n_p(\chi_p)}{2\gamma_p^3} \intl_{\frac{(\gamma_p+\gamma_e)^2}{4\gamma_p^2\gamma_e}}^{\frac{m_p}{\gamma_p m_e}}\td{\epsilon}\frac{n_{\rm ph}(\epsilon)}{\epsilon^2} \nonumber \\
	&\quad\times \intl_{\frac{(\gamma_p+\gamma_e)^2}{2\gamma_p\gamma_e}}^{2\gamma_p\epsilon}\td{\omega}\omega  \intl_{\frac{\gamma_p^2+\gamma_e^2}{2\gamma_p\gamma_e}}^{\omega-1}\td{E_{-}}\frac{W(\omega,E_{-})}{\sqrt{\frac{E_{-}^2}{c^2}-m_e^2c^2}}
	\label{eq:betheheitlerinj1},
\end{align}
where the initial factor $2$ accounts for electrons and positrons. The upper limit in the $\epsilon$-integral is a consequence of the Born approximation used in the cross section \citep{ka08}. The cross section $W(\omega,E_{-})$ is given in \cite{blumenthal70}, with $\omega$ and $E_{-}$ being the photon energy in units of $m_ec^2$ and the electron energy, respectively, in the proton rest frame. The integrals with respect to $\omega$ and $E_{-}$ depend solely on the electron momentum $\chi_e$, the proton momentum $\chi_p$ and the photon energy $\epsilon$. Therefore, we have tabulated these two integrals to save computation time in each time step.

Lastly, \g-\g\ pair production results in the injection term \citep{aan83}:

\begin{align}
	Q_{\rm \gamma\gamma}(\chi_e) &= 2\frac{3\sigma_Tc}{32}\intl_{\gamma_e}^{\infty}\td{\epsilon}\frac{n_{\rm ph}(\epsilon)}{\epsilon^3}\intl_{\frac{\epsilon}{4\gamma_e(\epsilon-\gamma_e)}}^{\infty}\td{\tilde{\epsilon}} \frac{n_{\rm ph}(\tilde{\epsilon})}{\tilde{\epsilon}^2} \nonumber \\
	&\quad\times \left[ \frac{4\epsilon^2}{\gamma_e(\epsilon-\gamma_e)}\ln{\left( \frac{4\gamma_e\tilde{\epsilon}(\epsilon-\gamma_e)}{\epsilon} \right)} - 8\epsilon\tilde{\epsilon} \right. \nonumber \\
	&\quad+ \left. \frac{2\epsilon^2(2\epsilon\tilde{\epsilon}-1)}{\gamma_e(\epsilon-\gamma_e)} - \left( 1-\frac{1}{\epsilon\tilde{\epsilon}} \right) \left( \frac{\epsilon^2}{\gamma_e(\epsilon-\gamma_e)} \right)^2 \right]
	\label{eq:gammagammainj1},
\end{align}
where, again, the leading factor $2$ accounts for electrons and positrons \citep{czbea21}. The photon distribution $n_{\rm ph}$ containing all internal and external photon fields, is evaluated at two normalized photon energies, namely $\epsilon$ and $\tilde{\epsilon}$ with the condition $\epsilon\gg\tilde{\epsilon}$.

%
%
\subsection{Acceleration terms}
We assume that pre-accelerated primary particles (protons and electrons) are injected throughout each slice. These may be accelerated in each slice at small turbulence regions or through gyroresonant interactions with magnetohydrodynamic waves. Such pre-acceleration zones are treated in codes such as \cite{weidingerspanier15} and \cite{chen+15} showing that power-law shaped particle distribution functions can be provided for the radiation zone. While we do not consider this pre-acceleration explicitly, we keep acceleration terms in Eq.~(\ref{eq:fpgen}) in order to provide a mild re-acceleration of the particles in the radiation zone. 

The momentum gain and loss rate in Eq.~(\ref{eq:fpgen}) is given as $\dot{\chi}_i = |\dot{\chi}_{i,{\rm loss}}| - \dot{\chi}_{\rm acc}$. The acceleration term contains Fermi-I acceleration, which is parameterized as

\begin{align}
     \dot{\chi}_{\rm acc} = \frac{\chi}{t_{\rm acc}}
     \label{eq:FermiIacc}.
\end{align}
The acceleration time scale is $t_{\rm acc} = \eta_{\rm acc}t_{\rm esc}$, that is a multiple of the escape time scale, where $\eta_{\rm acc}$ is a free parameter. 

Fermi-II acceleration is provided by the scattering of particles on magnetohydrodynamic waves. This results in momentum diffusion, described by the diffusion coefficient \citep{weidingerspanier15}

\begin{align}
     D(\chi) = \frac{\chi^2}{(2+a)t_{\rm acc}} 
     \label{eq:momdiffcoeff},
\end{align}
where we approximated the diffusion with hard-sphere scattering allowing for a momentum independent acceleration time scale. Following  \cite{weidingerspanier15}, the parameter $a=v_s^2/v_{A}^2$ is the ratio of the shock to the Alfv\`{e}n speed. For simplicity we set a fixed value of $a=10$ throughout the simulations. A critical assessment of this setting will be made elsewhere.

%
%
\subsection{Momentum loss terms}
The momentum loss term $\dot{\chi}_{i,{\rm loss}}$ depends on the particle species, as different loss processes are important for the different particles. We consider losses for protons through synchrotron, adiabatic, Bethe-Heitler and pion-production processes. Pions and muons lose momentum through synchrotron and adiabatic processes, while electrons lose momentum through synchrotron, IC and adiabatic processes. 

Synchrotron cooling depends on the magnetic field energy density $u_B = B^2/8\pi$ and the mass $m_i$ of the particle involved:

\begin{align}
     -\dot{\chi}_{i,{\rm syn}} = \frac{4c\sigma_T}{3m_ec^2}u_B \left( \frac{m_e}{m_i} \right)^3 \chi^2 
     \label{eq:cool-syn}.
\end{align}
The adiabatic term is adapted from \cite{zdziarski+14} as

\begin{align}
     -\dot{\chi}_{i,{\rm adi}} = \frac{3c\tan{(\eta_o/\Gamma_b)}}{R} \left( \gamma - \gamma^{-1} \right) 
     \label{eq:cool-adi}.
\end{align}
%
Protons lose energy also through Bethe-Heitler pair production, for which we use the semi-analytical result of \cite{chodorowski+92}:

\begin{align}
     -\dot{\chi}_{p,{\rm BH}} = \alpha_S r_e^2 c \frac{m_e}{m_p} \intl_2^{\infty} \td{\kappa} n_{\rm ph}\left( \frac{\kappa}{2\gamma} \right) \frac{\Phi(\kappa)}{\kappa^2}
     \label{eq:cool-BH},
\end{align}
where $\alpha_S\approx 1/137$ is the fine structure constant, $r_e$ the classical electron radius, $\kappa = 2\gamma\epsilon$, with $\epsilon = E_{ph}/m_ec^2$ being the normalized photon energy. We use the approximations to the cross-section integral $\Phi(\kappa)$ given in \cite{chodorowski+92}. 

At high energies, protons lose momentum predominantly through pion-production processes. We follow again the prescription of \cite{huemmer+10}. The loss rate 
is given by

\begin{align}
     -\dot{\chi}_{p,{\rm pion}} = \chi \suml_{\rm IT} M_{p}^{\rm IT} \Gamma^{\rm IT}(\gamma_p) K^{\rm IT} 
     \label{eq:cool-pion},
\end{align}
where the sum goes over all ITs that constitute the pion production cross section. In Eq.~(\ref{eq:cool-pion}), $M_p^{\rm IT}$ represents the multiplicity of daughter particles, while $K^{\rm IT}$ is the inelasticity of the process, and the interaction rate $\Gamma^{\rm IT}$ is given by

\begin{align}
     \Gamma^{\rm IT}(\gamma_p) = \intl_{\epsilon_{th}/2\gamma_p}^{\infty} \td{\epsilon} n_{\rm ph}(\epsilon) f^{\rm IT}(\gamma_p\epsilon)
     \label{eq:cool-pion-1}
\end{align}
%
We note again that $M_{p}^{\rm IT}$, $K^{\rm IT}$, and $f^{\rm IT}$ are tabulated allowing for a swift determination of the cooling term. Pion production might result in the conversion of a proton into a neutron. As we do not consider neutrons explicitly, we approximate this process as a continuous momentum loss process instead of an actual conversion using 

\begin{align}
     -\dot{\chi}_{p,{\rm neu}} = \chi \suml_{{\rm IT}, p\p\neq p} M_{p\p}^{\rm IT} \Gamma^{\rm IT}(\gamma_p)
     \label{eq:cool-neutron},
\end{align}
with the coefficients also provided by \cite{huemmer+10}. 

Charged pions and muons cool by synchrotron and adiabatic cooling, while electrons in addition exhibit IC cooling on the ambient photon field $n_{\rm ph}(\epsilon)$. The IC cooling term is \citep{boettcher+97}

\begin{align}
     -\dot{\chi}_{e,{\rm IC}} = c\pi r_e^2 \intl_{0}^{\infty}\td{\epsilon} n_{\rm ph}(\epsilon) \frac{G(\gamma_e \epsilon)}{\epsilon}
     \label{eq:cool-ic},
\end{align}
with

\begin{align}
     G(x) = \frac{8}{3}x\frac{1+5x}{(1+4x)^2} - \frac{4x}{1+4x}\left( \frac{2}{3}+\frac{1}{2x}+\frac{1}{8x^2} \right) \nonumber \\
     + \ln{(1+4x)}\left( 1+\frac{3}{x}+\frac{3}{4x^2}+\frac{\ln{(1+4x)}}{2x}-\frac{\ln{4x}}{x} \right) \nonumber \\
     - \frac{5}{2x} + \frac{1}{x} \suml_{1}^{\infty}\frac{(1+4x)^{-n}}{n^2} - \frac{\pi^2}{6x} - 2
     \label{eq:cool-ic-1}.
\end{align}
In the Thomson limit, that is $x\ll 1$, this can be Taylor-expanded as

\begin{align}
     \left. G(x)\right|_{x<0.2} \approx x^2 \left( \frac{32}{9}-\frac{112}{5}x+\frac{3136}{25}x^2 \right)
     \label{eq:cool-ic-2}.
\end{align}
The IC cooling term in this form requires isotropic photons. Therefore as mentioned before, the external photons are angle-averaged after the boosting into the comoving frame.

%
%
\subsection{Radiation terms} \label{sec:radterm}
In each time step, next to the Fokker-Planck equation for the particles, we also solve the radiative transfer equations for the photons. This ensures that the updated photon distribution can be used in the next time step for all particle-photon and photon-photon interactions. Along with the particle equilibria, an equilibrium solution for the photon distribution is found.

The radiative transport equation in the comoving frame of a slice is given by

\begin{align}
	\frac{\pd{n_{\rm ph}(\nu,t)}}{\pd{t}} = \frac{4\pi}{h\nu} j_{\nu}(t) - n_{\rm ph}(\nu,t) \left( \frac{1}{t_{\rm esc,ph}} + \frac{1}{t_{\rm abs}} \right) 
	\label{eq:radtrans}.
\end{align}
with the emissivity $j_{\nu}$ of all radiation processes, the photon escape time scale from a slice $t_{\rm esc,ph}(z)=4\Delta_{z}(z)/3c$, and the absorption time scale $t_{\rm abs}(z)$ due to synchrotron-self absorption and $\gamma$-$\gamma$ pair production. From the photon distribution, we can calculate the slice's spectral luminosity in the observer's frame

\begin{align}
	\nu\obs L\obs_{\nu\obs} = \delta_b^3 \frac{h\nu^2V_{\rm co}}{t_{\rm esc,ph}} n_{\rm ph}(\nu,t)
	\label{eq:speclum},
\end{align}
with the comoving volume of a slice $V_{\rm co} = \pi R(z)^2\Delta_{z}(z)$.

The synchrotron emissivity of a particle species with mass $m_i$ is given by \citep{bhk12}

\begin{align}
	j_{\nu,syn} = \frac{c\sigma_T u_B}{3\pi{\rm \Gamma}(4/3)} \left( \frac{m_e}{m_i} \right)^2 \nu^{1/3} \intl_{0}^{\infty}\td{\chi} n_i(\chi) \chi^2 \frac{e^{-\nu/\nu_c}}{\nu_c^{4/3}}
	\label{eq:synemis},
\end{align}
%
where ${\rm\Gamma}(x)$ is the Gamma-function, and 

\begin{align}
	\nu_c = \frac{3eB}{4\pi m_ic}\chi^2 
	\label{eq:critsynfreq}.
\end{align}

Electrons also undergo IC emission with the emissivity given by \citep{db14,dm09}

\begin{align}
    j_{\epsilon_s,iso} &= A_i \epsilon_{s} \intl_{-1}^{1}\td{\mu} \frac{1-\cos{\Psi}}{(\Gamma_b(1+\beta_b\mu))^2} \intl_{\chi_{\rm min}}^{\infty} \td{\chi} \frac{\chi n_e(\chi)}{\gamma}\Sigma_c(\gamma,\mu)
    \label{eq:jeciso},
\end{align}
which holds for isotropic photon fields in the galaxy frame. Here, $\epsilon_s$ is the normalized scattered photon energy,

\begin{align}
    \cos{\Psi} &= \mu\mu_s+\sqrt{1-\mu^2}\sqrt{1-\mu_s^2} \label{eq:cospsi} \\
    \mu_s &= \frac{\cos{\theta_{\rm obs}}-\beta_b}{1-\beta_b\cos{\theta_{\rm obs}}} \label{eq:ecmus} \\
    \chi_{\rm min} &= \frac{\epsilon_s}{2} \left( 1+\sqrt{1+\frac{2\Gamma_b(1+\beta_b\mu)}{\hat{\Theta}_i\epsilon_s(1-\cos{\Psi})}} \right) \label{eq:ecchimin} \\
    \Sigma_c(\gamma,\mu) &= \frac{8\sigma_T}{3\gamma \epsilon_0} \left( y+\frac{1}{y}-\frac{2\epsilon_s}{y\gamma\epsilon_0} + \left( \frac{\epsilon_s}{y\gamma\epsilon_0} \right)^2 \right) \nonumber \\
    &\quad \times \SF{\epsilon_s}{\frac{\epsilon_0}{2\gamma}}{\frac{2\gamma\epsilon_0}{1+2\epsilon_0}} \label{eq:eccs} \\
    y &= 1-\frac{\epsilon_s}{\gamma} \label{eq:ecy} \\
    \epsilon_0 &= \gamma\frac{\hat{\Theta}_i}{\Gamma_b(1+\beta_b\mu)} (1-\cos{\Psi}) \label{eq:eceps0} \\
    \hat{\Theta}_i &= \frac{2.7 k_B \hat{T}_i}{m_ec^2} \label{eq:ectheta}
\end{align}
where the temperatures of the external fields are free parameters, except for the CMB with $\hat{T}_{\rm CMB}=2.72(1+z_{\rm red})\,$K. The constants $A_i$ depend on the photon field:

\begin{align}
    A_{\rm BLR} &= \frac{h\hat{L}_{\rm BLR}(z)}{4\pi \hat{R}_{\rm BLR}^2 (2.7k_B \hat{T}_{\rm BLR})} \label{eq:Ablr} \\
    A_{\rm DT} &= \frac{h\hat{L}_{\rm DT}(z)}{4\pi \hat{R}_{\rm DT}^2 (2.7k_B \hat{T}_{\rm DT})} \label{eq:Adt} \\
    A_{\rm CMB} &= \frac{8\pi^5m_ec^4}{15} \left[ (1+z_{\rm red})\Theta_{\rm CMB}^{{\rm obs}} \right]^3 \label{eq:Acmb}, 
\end{align}
for photons from the BLR, the DT, and the CMB, respectively. We stress that $\Theta\obs_{\rm CMB}$ is defined in the observer's frame at redshift $z_{\rm red}=0$. 
Scattering AD photons, the IC emissivity becomes \citep{db14}

\begin{align}
    j_{\epsilon_s,AD} &= A_{\rm AD} \epsilon_{s} \intl_{\mu_{min}}^{\mu_{max}}\td{\mu_d} \frac{(1-\cos{\Psi})\,\left[ \left( \frac{1+\beta_b\mu_d}{\beta_b+\mu_d} \right)^2-1 \right]^6}{\tilde{\Theta}_{\rm AD}[\Gamma_b(1+\beta_b\mu_d)]^3} \nonumber \\
    &\quad\times \intl_{\chi_{\rm min}}^{\infty} \td{\chi} \frac{\chi n_e(\chi)}{\gamma}\Sigma_c(\gamma,\mu_d)
    \label{eq:jecdisk},
\end{align}
with

\begin{align}
    A_{\rm AD} &= \frac{3hGM_{0}\dot{m}}{8\pi m_ec^2z^3} \label{eq:Aacc} \\
    \tilde{\Theta}_{\rm AD} &= \frac{2.7 k_B \hat{T}(\hat{r}_{\rm AD})}{m_ec^2} \label{eq:thetaacc} \\
    \mu_d &= \frac{\frac{z}{\sqrt{z^2+\hat{r}_{\rm AD}^2}}-\beta_b}{1-\beta_b\frac{z}{\sqrt{z^2+\hat{r}_{\rm AD}^2}}} \label{eq:mudisk} ,
\end{align}
$\mu_{min}=\mu_d(\hat{R}_{\rm AD,min})$, and $\mu_{max}=\mu_d(\hat{R}_{\rm AD,max})$, cf. Sec.~\ref{sec:extfield}. 
Most relations of Eqs.~(\ref{eq:cospsi})-(\ref{eq:eceps0}) hold, provided $\mu=\mu_d$ and $\hat{\Theta}_i=\tilde{\Theta}_{\rm AD}$. We reemphasize the radial dependence of the AD parameters. 
In Eqs.~(\ref{eq:jeciso}) and (\ref{eq:jecdisk}), we use a $\delta$-function approximation to the energy distribution of the external photon fields using the peak of the thermal distributions at $E=2.7k_BT$, while we consider the full angle-dependence of the beaming pattern \citep{db14}.

We also consider SSC emission for electrons, which is calculated according to \citep{db14} 

\begin{align}
    j_{\epsilon_s,ssc} &= \frac{h\epsilon_s}{4\pi} \intl_{0}^{\infty}\td{\chi} \frac{\chi}{\gamma}n_e(\chi)  \intl_{0}^{\infty}\td{\epsilon} n_{\rm syn}(\epsilon) G(\epsilon_s,\epsilon,\gamma) \label{eq:jssc},
\end{align}
where the synchrotron photon distribution is calculated with Eqs.~(\ref{eq:radtrans}) and (\ref{eq:synemis}), while

\begin{align}
    G(\epsilon_s,\epsilon,\gamma) &= \begin{cases}
            \frac{8\sigma_Tc}{6\epsilon\gamma^4} \left( \frac{4\epsilon_s\gamma^2}{\epsilon}-1 \right) & \frac{\epsilon}{4\gamma^2} < \epsilon_s \leq \epsilon \\
            \frac{16\sigma_Tc}{3\gamma^2\epsilon} G_{q_s} & \epsilon < \epsilon_s \leq \frac{4\epsilon\gamma^2}{1+4\epsilon\gamma}
        \end{cases} \label{eq:Gssc} \\
    G_{q_s} &= \left[ 2q_s\ln{q_s}+(1+2q_s)(1-q_s) \right. \nonumber \\
    &\left.+(1-q_s)\frac{(4\epsilon\gamma q_s)^2}{2(1+4\epsilon\gamma q_s)} \right] \label{eq:Gqssc} \\
    q_s &= \frac{\epsilon_s}{4\epsilon\gamma(\gamma-\epsilon_s)} \label{eq:qssc}.
\end{align}

Neutral pions decay directly into photons. The resulting photon emissivity is given by \citep[e.g.][their Eq.~(3.100)]{bhk12}

\begin{align}
    j_{\nu,\pi^0} = 2\frac{h^2\nu}{4\pi}\intl_{\chi_{min}}^{\infty}\td{\chi} \frac{Q_{\pi^0}(\chi)}{\sqrt{\chi^2+1}} 
    \label{eq:jpizero},
\end{align}
with 
the lower limit of this integral

\begin{align}
    \chi_{min} = \sqrt{\left( \frac{h\nu}{E_{\pi^0}} + \frac{E_{\pi^0}}{4h\nu} \right)^2 - 1}
    \label{eq:pizeromin}.
\end{align}

Lastly, in order to complete Eq.~(\ref{eq:radtrans}), the absorption time scale is required:

\begin{align}
	t_{\rm abs} = \frac{\Delta_{z}}{c(\tau_{\rm SSA}(\nu)+\tau_{\rm \gamma\gamma}(\epsilon))}
	\label{eq:abstime},
\end{align}
where $\tau_{\rm SSA}$ and $\tau_{\rm \gamma\gamma}$ are the synchrotron-self-absorption and pair production opacities, respectively. The synchrotron-self absorption opacity for a charged particle of mass $m_i$ is \citep{dm09}

\begin{align}
    \tau_{\rm SSA}(\nu) = -\frac{\Delta_{z}}{8\pi m_i \nu^2}\intl_0^{\infty}\td{\chi} P_{\nu,syn}(\chi)\chi^2 \frac{\pd}{\pd{\chi}} \left[ \frac{n_i(\chi)}{\chi^2} \right]
    \label{eq:taussa}.
\end{align}

The pair production opacity is given by \citep{dm09} 

\begin{align}
    \tau_{\gamma\gamma}(\epsilon) = \frac{8\Delta_{z}\sigma_T}{3\epsilon^2} \intl_{\epsilon^{-1}}^{\infty}\td{\tilde{\epsilon}} \frac{n_{\rm ph}(\tilde{\epsilon})}{\tilde{\epsilon}^2} \bar{\varphi}(s)
    \label{eq:taugg}
\end{align}
with $s = \epsilon\tilde{\epsilon}$, and the cross section

\begin{align}
    \bar{\varphi}(s-1\ll 1) &\approx \frac{4}{3}(s-1)^{3/2} + \frac{6}{5} (s-1)^{5/2} \nonumber \\
    &\quad - \frac{253}{70}(s-1)^{7/2} \label{eq:tauggphi1} \\
    \bar{\varphi}(s\gg 1) &\approx 2s(\ln{4s}-2)+\ln{4s}\, (\ln{4s}-2) - \frac{\pi^2-9}{3} \nonumber \\
    &\quad + \frac{\ln{4s}+9/8}{s} \label{eq:tauggphi2}.
\end{align}
We split the approximations of the cross section at $s=1.4$ \citep[cf.][Fig. 10.2]{dm09}. Equation~(\ref{eq:taugg}) requires isotropic photon fields, which is fulfilled for internal photon distributions. The external photon fields are -- similar to the proton-photon interactions -- angle-averaged after beaming into the comoving frame.

We note that for both synchrotron-self absorption and internal pair production, we only consider the slice where the emission is produced. This is a simplification, which we intend to improve in a subsequent paper.

%
%
\subsection{Neutrinos} \label{sec:neu}
Neutrinos are created through the decay of pions and muons. In the former case, muon neutrinos are created, while in the latter case both muon and electron neutrinos are produced. The production rate of muons from pion decay $Q_{\mu^{\pm}}$, Eq.~(\ref{eq:muoninjection}), directly provides us with the production rate of muon neutrinos:

\begin{align}
	Q_{\nu_{\mu}}^{\pi}(E_{\nu_{\mu}}) = \frac{1}{m_{\mu}c^2} \intl_{\frac{E_{\nu_{\mu}}}{m_{\pi}c^2(1-r_M)}}^{\infty} \frac{\td{\gamma_{\pi}}}{\gamma_{\pi}} \frac{Q_{\mu^{\pm}}(\gamma_{\pi})}{1-r_M}
	\label{eq:q_neu_mu_pion},
\end{align}
where $r_M = (m_{\mu}/m_{\pi})^2$. 

For the muon decay, Eqs.~(\ref{eq:muonpdecay}) and (\ref{eq:muonmdecay}), we follow the description of \cite{barr+88} and \cite{gaisser90} with the neutrino production rate for both types given as

\begin{align}
	Q_{\nu_i}^{\mu}(E_{\nu_i}) &= \frac{1}{m_{\mu}c^2} \intl_{E_{\nu_i}}^{\infty} \td{E_{\mu}} Q_{e}^{\mu}(E_{\mu}) \frac{\td{n}}{\td{E_{\nu_i}}} \nonumber \\
	&= \frac{1}{m_{\mu}c^2} \intl_0^1 \td{y} \frac{Q_{e}^{\mu}(E_{\nu_i}/y)}{y} \frac{\td{n}}{\td{y}}
	\label{eq:q_neu_mu_e},
\end{align}
with $y=E_{\nu_i}/E_{\mu}$, and the muon decay rate $Q_{e}^{\mu}=n_{\mu^i}(\chi)/(\gamma t\p_{\mu^i,{\rm decay}})$. Both $Q_{e}^{\mu}$ and $Q_{\mu^{\pm}}$ are in units of per volume per time, while the neutrino production rates require per volume per time per energy, explaining the normalization of the integrals to the restmass energy of the muon. The function $\td{n}/\td{m}$ describes the neutrino production rate in the laboratory frame and is approximately \citep{gaisser90} 

\begin{align}
	\frac{\td{n}}{\td{y}} \approx g_0(y) + g_1(y)
	\label{eq:neu_prod_rate1}.
\end{align}
The functions $g_0$ and $g_1$ depend on the neutrino type and are for muon neutrinos

\begin{align}
	g_0(y) &= 5/3 - 3y^2 + 4y^3/3 \label{eq:neu_mu_g0} \\
	g_1(y) &= 1/3 - 3y^2 + 8y^3/3 \label{eq:neu_mu_g1},
\end{align}
while for electron neutrinos

\begin{align}
	g_0(y) &= 2 - 6y^2 + 4y^3 \label{eq:neu_e_g0} \\
	g_1(y) &= -2 + 12y - 18y^2 + 8y^3 \label{eq:neu_e_g1}.
\end{align}

Owing to the oscillation of neutrinos, the observer's frame neutrino power $E\obs_{\nu_i}L\obs_{E\obs_{\nu_i}}=\delta_b^3(E_{\nu_i}/(1+\zred))^2 V_{co}Q_{\nu_i}$ is equally distributed over electron, muon and tau neutrinos. Therefore, $Q_{\nu_i} = (Q_{\nu_{\mu}}^{\pi}+Q_{\nu_{\mu}}^{\mu}+Q_{\nu_e}^{\mu})/3$. In the given framework, we have not distinguished between neutrinos and their anti-particles.

%
%
\section{Results} \label{sec:results}
\begin{table*}
\caption{Overview of the free parameters. Baseline parameters are used in simulation 01, while parameters listed under Variation 1 and 2 are used in simulations given in parentheses. The value of the AD Eddington ratio defines the cases A and B of all simulation.}
\begin{tabular}{lcl|ccc}
Definition		& \multicolumn{2}{c|}{Symbol} 		& Baseline	& Variation 1	& Variation 2 \\
\hline
Redshift    & $z_{\rm red}$  &   & $0.5$  & -  & -  \\
Black hole mass    & $M_{0}$  & [$10^8M_{\odot}$]  & $3.0$  & -  & -  \\
AD Eddington ratio    & $l_{\rm edd}$  &   & -  & $10^{-1}$ (\textcolor{red}{\textbf{A}}) & $10^{-3}$ (\textcolor{blue}{\textbf{B}})  \\
BLR temperature    & $T_{\rm BLR}$  & [K]  & $10^4$  & -  & -  \\
DT temperature    & $T_{\rm DT}$  & [K]  & $5\E{2}$  & -  & -  \\
Jet length    & $z_{\rm term}$  & [pc]  & $100.0$  & -  & -  \\
Length of acceleration region    & $z_{\rm acc}$  & [pc]  & $1.0$  & $0.1$ (\textbf{02})  & $10.0$ (\textbf{03})  \\
Maximum bulk Lorentz factor    & $\Gamma_{b,{\rm max}}$  &   & $30$  & $15$ (\textbf{04})  & $50$ (\textbf{05})  \\
Jet viewing angle    & $\theta_{\rm obs}$  & [deg]  & $1.9$  & $3.8$ (\textbf{04})  & $1.1$ (\textbf{05})  \\
Multiple of jet opening angle    & $\eta_{o}$  &    & $0.26$  & -  & -  \\
Multiple of initial jet radius    & $\eta_{R}$  &   & $10$  & -  & -  \\
Mulitple of escape time scale    & $\eta_{\rm esc}$  &   & $10$  & -  & -  \\
Multiple of acceleration time scale    & $\eta_{\rm acc}$  &   & $10$  & -  & -  \\
Initial magnetic field    & $B(z_0)$  & [G]  & $50$  & $30$ (\textbf{06})  & $100$ (\textbf{07})  \\
Multiple of injection power    & $f_{\rm inj}$  &   & $3\E{-6}$  & $3\E{-7}$ (\textbf{08})  & -  \\
Initial proton to electron ratio    & $\kappa_{pe}$  &   & $1.0$  & $0.1$ (\textbf{09})  & -  \\
Minimum proton Lorentz factor    & $\gamma_{p,1}$  &   & $2$  & -  & -  \\
Maximum proton Lorentz factor    & $\gamma_{p,2}$  &   & $2\E{8}$  & $2\E{7}$ (\textbf{10})  & $5\E{8}$ (\textbf{11})  \\
Proton spectral index    & $p_p$  &   & $2.5$  & $2.1$ (\textbf{12})  & $3.0$ (\textbf{13})  \\
Minimum electron Lorentz factor    & $\gamma_{e,1}$  &   & $1\E{2}$  & -  & -  \\
Maximum electron Lorentz factor    & $\gamma_{e,2}$  &   & $1\E{5}$  & $1\E{4}$ (\textbf{14})  & $1\E{6}$ (\textbf{15})  \\
Electron spectral index    & $p_e$  &   & $2.5$  & $2.1$ (\textbf{16}) & $3.0$ (\textbf{17}) \\
\end{tabular}
\label{tab:freepara}
\end{table*}
In Table~\ref{tab:freepara} we provide an overview over all free parameters. The baseline simulation 01 is described in detail in section~\ref{sec:sim01}. From these parameters we vary other parameters one at a time for the parameter study in section~\ref{sec:parastudy}. For each simulation, we produce two SEDs -- "A" and "B" -- where the only difference is the value of the AD Eddington ratio. As the BLR and the DT also depend on this value according to Eqs.~(\ref{eq:RBLR}) to (\ref{eq:LDTz}), this is going to have significant consequences on the results. 

%
%
\subsection{Baseline model} \label{sec:sim01}
We use simulation 01 as our baseline to describe in detail the capabilities of \textit{ExHaLe-jet}. The results of the other simulations are briefly summarized in section~\ref{sec:parastudy}.

%
%
\subsubsection{Photon spectra} \label{sec:sim01phot}
\begin{figure*}
\centering 
\includegraphics[width=0.95\textwidth]{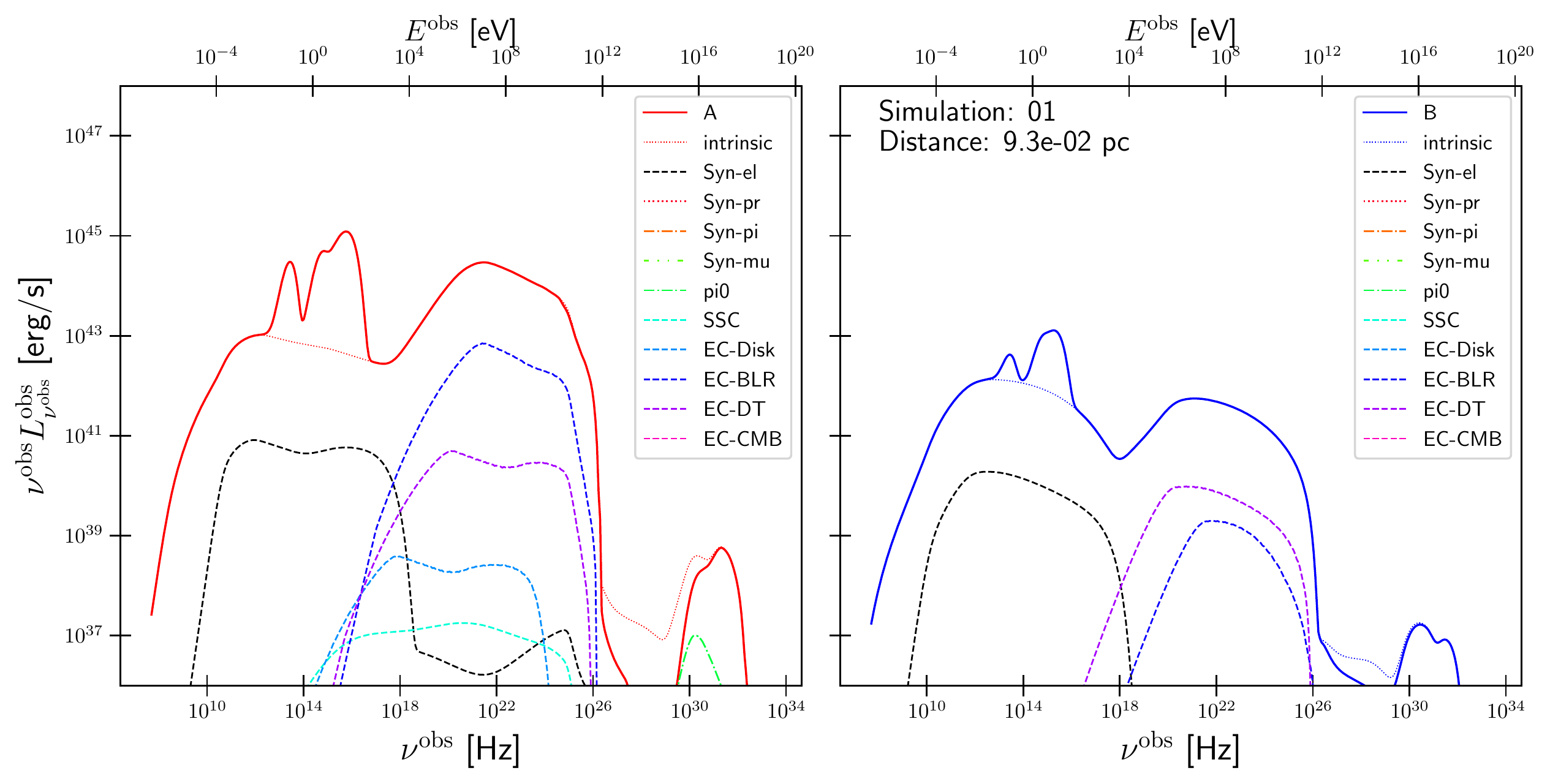}
\caption{Total photon spectrum in the observer's frame and the individual contributions for the slice at the indicated distance for simulation 01 A (left) and B (right). The thin dotted line shows the total intrinsic spectrum, while the thick solid line includes the external photon fields as well as the (external) absorption at \g-ray frequencies. The remaining lines show the contributions of the different radiation processes as labeled. The proton-, charged pion- and muon-synchrotron spectra are below the shown luminosity scale.
}
\label{fig:run01_specexam}
\end{figure*} 
\begin{figure*}
\centering 
\includegraphics[width=0.95\textwidth]{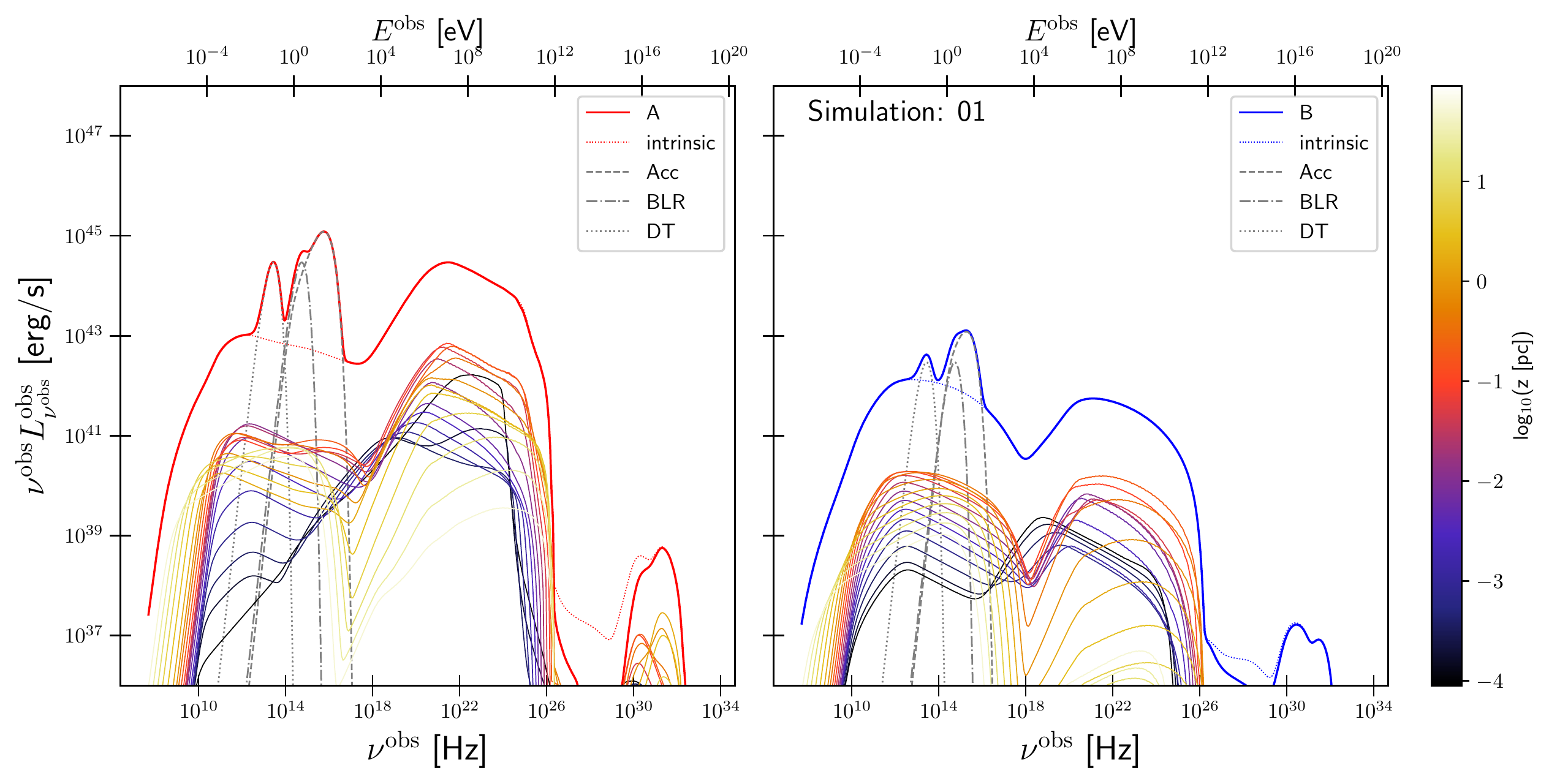}
\caption{Total photon spectrum in the observer's frame and its distance dependence (color code) for simulation 01 A (left) and B (right). The thin dotted line shows the total intrinsic (i.e. no absorption outside the slice) spectrum, while the thick solid line includes the external photon fields (gray) as well as the (external) absorption at \g-ray frequencies. The thin colored lines show the intrinsic spectrum of every tenth slice.
}
\label{fig:run01_specdist}
\end{figure*} 
The multiwavelength photon spectra of the baseline simulation are shown in Fig.~\ref{fig:run01_specexam} for both cases of the accretion disk luminosity. It is obvious that the total spectra depend strongly on the external photon fields, as the ratio of the \g-ray peak luminosity to the electron-synchrotron peak luminosity is larger than unity for case A, while it is smaller than unity for case B. This is a consequence of a significant decrease in the \g-ray flux, while the electron-synchrotron flux only decreases mildly from case A to B. 

The individual spectra for the example slice at about $0.1\,$pc from the black hole indicate that the \g-ray flux is dominated by IC emission on the BLR and DT photon fields. The dependence of the BLR and DT on the accretion disk luminosity explains the change in the relative strength of the IC/BLR and IC/DT. In case A, the radius of the BLR (DT) is $0.06\,$pc ($1.6\,$pc), while in case B it is $0.006\,$pc ($0.16\,$pc)\footnote{We recall that the BLR and DT luminosities turn into a power-law dependence on $z$ at their respective outer radii according to Eqs.~(\ref{eq:LBLRz}) and (\ref{eq:LDTz}). This explains why the IC/BLR process can still be hugely dominating over (be comparable with) the IC/DT process even though the considered emission region is (far) outside the BLR radius in case A (B).}. 
The higher external photon density in case A compared to case B implies a faster electron cooling in case A because of IC cooling than in case B, as discussed in section~\ref{sec:sim01par}. 

A second influence of the external fields is visible through the different degree of external absorption at TeV energies. In Fig.~\ref{fig:run01_specexam} the thin dotted line marks the intrinsic spectrum, that is the sum of the slices boosted into the observer's frame but discarding any absorption outside of the jet, while the thick solid line marks the jet emission after considering absorption in the BLR and DT. While the emission is attenuated between $\sim 1$\,TeV and a few PeV by up to a few orders of magnitude in case A, there is barely a difference in case B.

At photon energies above one PeV, a third bump emerges, which is due to the decay of neutral pions. The photon densities below meV energies (required to absorb the \g-ray photons of the neutral pions) are not sufficient to absorb these \g\ rays entirely. This also means that the emission of the neutral pions does not take part significantly in the development of the pair cascade. The neutral pion bump is most likely not observable at Earth given the cosmological absorption through the EBL and the CMB, which is not considered here. Nevertheless, the external photon fields also play a role in the luminosity of the neutral pion bump as the luminosity in case A is about two orders of magnitude higher than in case B. In turn, the pion production largely depends on the external photon fields. This is also evident from the neutrino output, as discussed in section~\ref{sec:sim01neu}.

Notably absent from Fig.~\ref{fig:run01_specexam} are the synchrotron emission of protons, charged pions and muons. Their densities are too low to produce meaningful radiative components. On the other hand, the electron-synchrotron emission extends well into the \g-ray regime. This extension is a consequence of the highly energetic secondaries injected in each slice.
Interestingly, SSC emission is also irrelevant, while IC/CMB starts to become important at larger distances $z$ (cf. Fig.~\ref{fig:run01_speccompdist}). 

The non-trivial evolution of the photon spectra with distance $z$ is shown in Fig.~\ref{fig:run01_specdist}. 
In both cases, A and B, the electron-synchrotron component increases gradually until about $0.1\,$pc from the black hole and remains relatively steady (even more in case B than in case A) until it starts to decrease about $10\,$pc from the jet.

On the contrary, the \g-ray component is dominated initially in both cases by IC on disk photons. Interestingly, in case A the IC/AD flux is initially very strong and decreases rapidly. In case B the IC/AD flux decreases, too, but at a much lower flux level. 
This is a consequence of Eq.~(\ref{eq:ADrmax}), as the outer disk radius influences strongly the region-of-influence of the AD on the IC process given that the outer regions of the disk exhibit a different beaming pattern than the inner disk parts. As Eq.~(\ref{eq:ADrmax}) depends on the Eddington ratio $l_{\rm edd}$, the disk is less wide in case B than in case A. With increasing distances from the black hole, the IC process in the jet becomes first dominated by BLR photons and then by DT photons. This obviously depends on the respective radii as discussed above. The maximum IC luminosity is attained in the range from $0.1$ to $1\,$pc from the black hole irrespective of the case. The neutral pion bump evolution shows two peaks. In case A, the first peak is attained at $0.1\,$pc from the black hole and is located at slightly lower energies than the second bump, which is attained at about $1\,$pc from the black hole. Apparently, the BLR and DT photon fields with their different energy distributions interact with the protons at different distances from the black hole producing pions of different energies. In case B, the first peak is seemingly produced with AD photons, while the second peak is probably due to DT photons. 

While especially at \g-ray energies the resolution power will not be sufficient to resolve most jets -- with the noteworthy exception of Centaurus~A \citep{hess20} -- it is still an important question, where the \g\ rays are produced within the jet given the potential absorption processes. Within our model, the \g\ rays do not emerge from a single region, but are produced within $10\,$pc ($1\,$pc) in case A (B). While this may be a result of our steady injection spectrum along the jet, it nonetheless emphasizes that \g\ rays can be produced on very different scales.

%
%
\subsubsection{Evolution of the particle distributions} \label{sec:sim01par}
\begin{figure*}
\centering 
\includegraphics[width=0.90\textwidth]{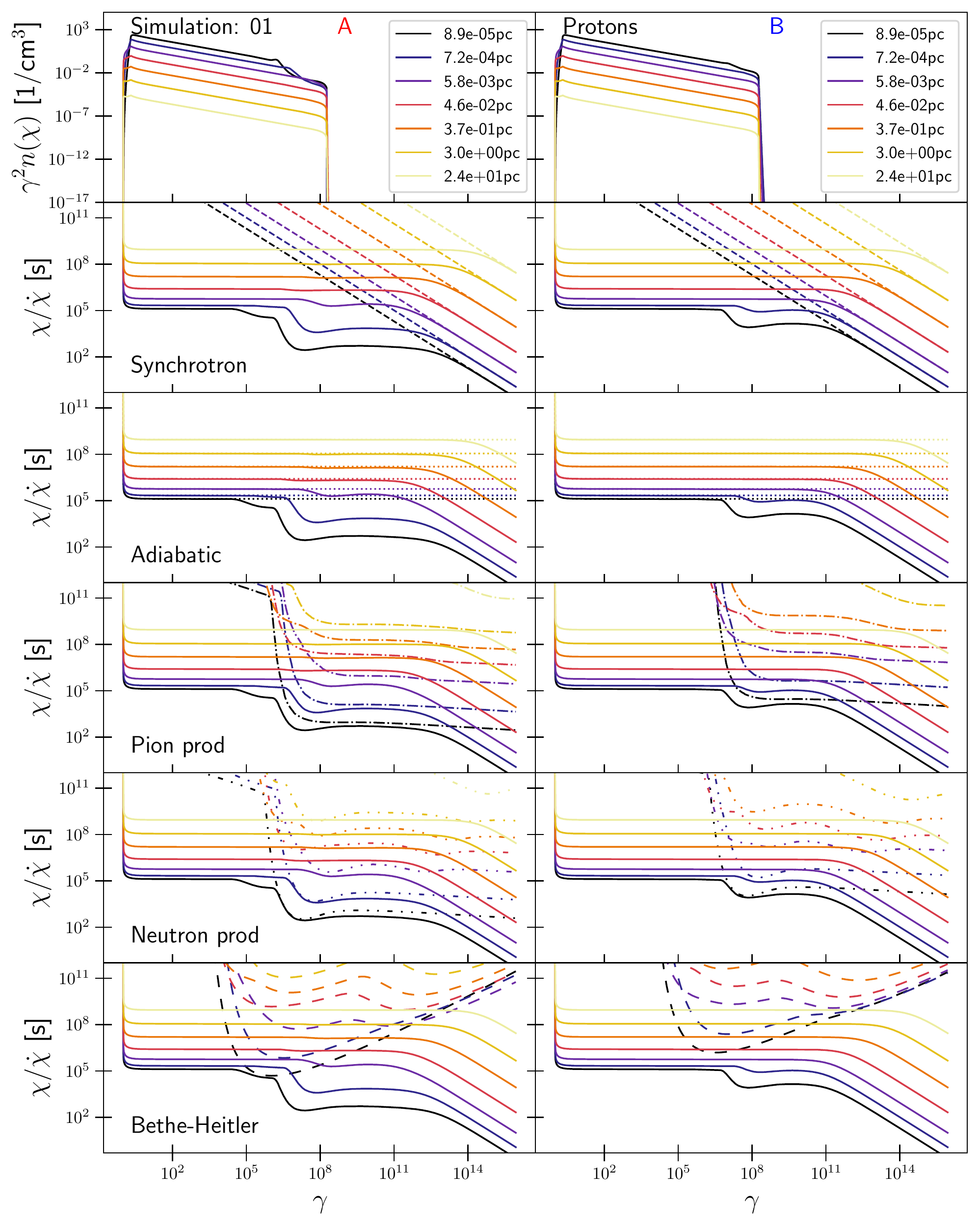}
\caption{Proton distribution function (top row) and cooling time scales as a function of Lorentz factor $\gamma$ and distance $z$ (color code) as labeled for simulation 01 A (left) and B (right). For the cooling time scales, the solid lines markes the total cooling time scale, while the lines with a different style mark the individual process as labeled. The curves are in the comoving frame of each slice.
}
\label{fig:run01_partpr}
\end{figure*}
\begin{figure*}
\centering 
\includegraphics[width=0.90\textwidth]{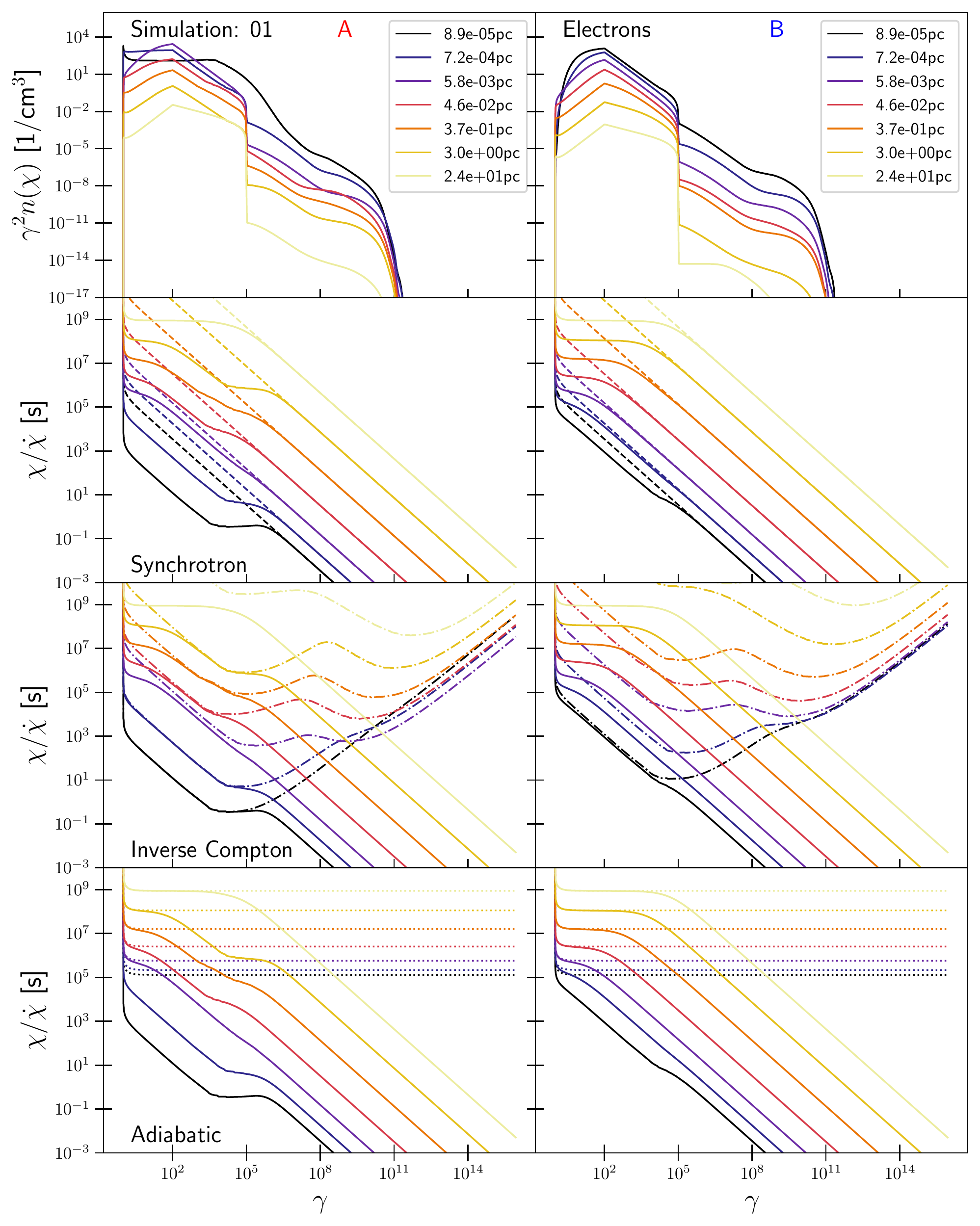}
\caption{Same as Fig.~\ref{fig:run01_partpr} but for electrons.
}
\label{fig:run01_partel}
\end{figure*}
%
%
%
\begin{figure*}
\centering 
\includegraphics[width=0.90\textwidth]{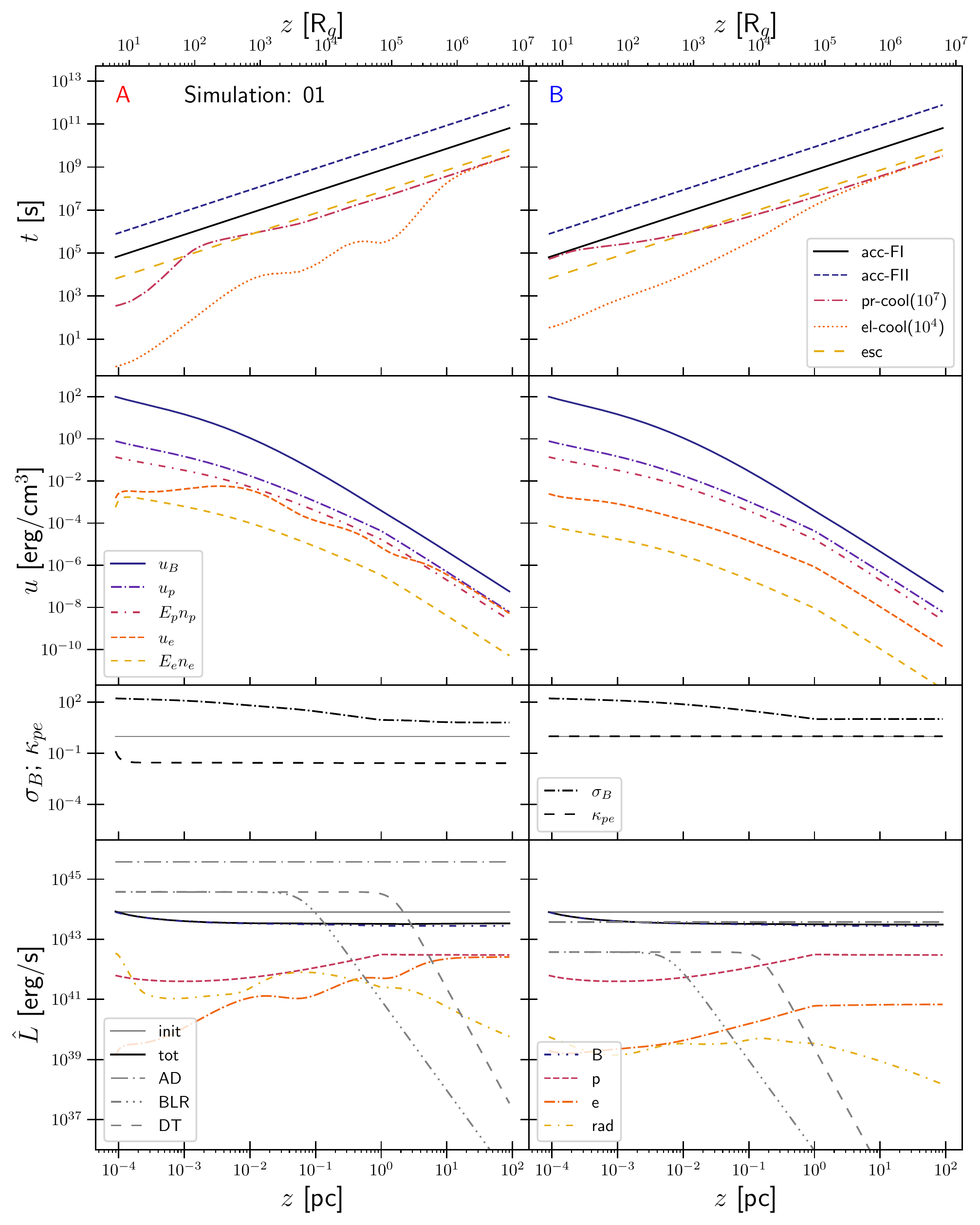}
\caption{Jet parameters as a function of distance $z$ along the jet for simulation 01 A (left) and B (right). The legend in each row corresponds to both panels.
\textit{Top row:} Evolution of Fermi-I-, and Fermi-II-acceleration, proton and electron (dotted), and escape time scales as labeled. The cooling time scale is taken exemplary at proton and electron Lorentz factors of $10^7$ and $10^4$, respectively. The time scales are derived in the comoving frame of each slice. 
\textit{Second row:} Magnetic energy density, energy densities $u_i$ of protons and electrons, and the rest-mass energy densities $E_in_i$ of protons and electrons in the comoving frame of each slice.
\textit{Third row:} Magnetization $\sigma_B$ and proton-to-electron ration $\kappa_{pe}$ in the comoving frame of each slice. The gray solid line marks unity.
\textit{Fourth row:} Magnetic, proton, electron, and radiative luminosities. Also shown is the total of these luminosities for each slice, while the initial value (``init'') and the accretion disk are given as reference. The evolution of the BLR and DT luminosities are also provided. All luminosities are in the host galaxy's frame.
}
\label{fig:run01_paraall}
\end{figure*}
The evolution of the proton and electron distributions as a function of distance $z$ is shown in Figs.~\ref{fig:run01_partpr} and \ref{fig:run01_partel}, respectively. We also show in the same figures the cooling time scales highlighting the different processes shaping the particle distributions.

The proton distributions do not differ strongly between case A and B except at Lorentz factors above $10^6$ for small distances $z$. Here, a dip is visible in case A owing to stronger cooling through pion and Bethe-Heitler pair production. Clearly, the stronger external photon fields are responsible for the enhanced cooling in case A compared to case B. 
At lower Lorentz factors, the cooling is dominated by adiabatic losses, while synchrotron cooling is negligible for the proton distribution in this simulation.

The electron distributions in Fig.~\ref{fig:run01_partel} show remarkable features. While the primary injection spectrum between Lorentz factors $10^2$ and $10^5$ is visible, the secondary particles play a major role in shaping the final distribution (see also App.~\ref{app:sec} and Fig.~\ref{fig:run01_partseco}). In turn, the electron distributions extend to very high Lorentz factors beyond $10^{11}$. The imprint of the different external photon fields between case A and B is notable in the particle distributions and the cooling time scales. Case A exhibits a higher number of secondaries which at low distances $z$ also influence the distribution at Lorentz factors below $\gamma_{e,2}$. Such a significant influence is absent in case B. The cooling time scale in case A is dominated by IC processes at Lorentz factors below $10^5$. Above this threshold, the Klein-Nishina effect significantly reduces the IC efficiency, and the synchrotron process starts to dominate. However, for greater distances $z$, the IC strength is also reduced compared to the synchrotron, and beyond a few pc -- corresponding to the DT radius -- the IC influence becomes negligible. In fact, with increasing distance $z$ adiabatic cooling becomes important at lower and medium Lorentz factors. In case B, similar statements can be made with the difference that the IC process is less severe due to the weaker external fields. In turn, the overall cooling strength is also weaker in this case.


The cooling time scales of protons and electrons (taken exemplary at Lorentz factors of $10^7$ and $10^4$, respectively) are compared to acceleration and escape time scales in the top row of Fig.~\ref{fig:run01_paraall}. It is evident that the electron cooling time scale is always below the escape and acceleration time scales indicating that the electrons are in the fast cooling regime at all $z$ for both cases A and B. The protons in case A -- at least at this Lorentz factor -- initially cool faster than they escape, which then changes between distances of $10^{-3}$ and $0.01\,$pc, beyond which the cooling is again faster than the escape. In case B, the protons initially cool much slower than they escape. In fact at low distances, the cooling time scale is comparable to the Fermi-I acceleration time scale. Only at distances beyond $0.01\,$pc is the cooling faster than the escape. As the escape time scale depends on the length $\Delta_{z}(z)$ of a slice, its increase with distance is evident. Similarly, the Fermi-I and II acceleration time scales depends directly on the escape time scale. 

%
%
\subsubsection{Jet evolution} \label{sec:sim01jet}
%
%
%
In the second row of Fig.~\ref{fig:run01_paraall}, we show the energy densities of the magnetic field and the particles. The magnetic energy density is $u_B(z)=B(z)^2/8\pi$, while the total and rest-mass energy densities for protons and electrons are 

\begin{align}
    u_i(z) &= m_ic^2 \intl_0^{\infty} \gamma n_i(\chi,z) \td{\chi} \label{eq:partendens} \\
    E_in_i(z) &= m_ic^2 \intl_0^{\infty} n_i(\chi,z) \td{\chi} \label{eq:partrestendens},
\end{align}
respectively. 

Both disk cases are actually similar for most constituents. The magnetic energy density dominates at all distances, and the proton values dominate over the electron values except at large distances in case A. 
The main difference between the cases is the energy density in electrons, as the (initial) stronger cooling in case A results in a lower energy density of the electrons compared to case B. Only at larger distances, when the IC cooling becomes less severe, do the cases match again. 

The dominance of the magnetic energy density at all distances also implies that the magnetization $\sigma_B(z)$, Eq.~(\ref{eq:magnetization}), is larger than unity on all scales as shown in the third row of Fig.~\ref{fig:run01_paraall}. While $\sigma_B(z)$ decreases in the parabolic section of the jet, it is constant in the conical section, as expected. 
The high magnetization implies that our jet would energize the particles via magnetic reconnection on all scales. However, as we only intend to perform initial tests here, this is not a major concern. Different parameter sets result in lower magnetizations on large scales (cf. Tab.~\ref{tab:magkappa} in App.~\ref{app:figs}).

Additionally in the third row of Fig.~\ref{fig:run01_paraall} we show the evolution of $\kappa_{pe}$. In both cases, the initial value is unity and there is no significant change to that in case B. In case A, close to the AD \g-\g\ pair production is strong and significantly decreases $\kappa_{pe}$. It remains constant at larger distances.

The fourth row of Fig.~\ref{fig:run01_paraall} shows the various jet luminosities in the host galaxy's frame. For the magnetic field and the particles, the luminosity is calculated as

\begin{align}
    \hat{L}_i(z) = \pi R(z)^2 \Gamma_b(z)^2 c u_i(z) 
    \label{eq:luminosity1},
\end{align}
while the total radiative luminosity is

\begin{align}
    \hat{L}_{\rm rad}(z) = \frac{\Gamma_b(z)^2}{\delta(z)^3} \intl_0^{\infty} L\obs_{\nu\obs}(z) \td{\nu\obs} 
    \label{eq:luminosity2}.
\end{align}
These are compared to the injected (or initial) luminosity at the base of the jet:

\begin{align}
    \hat{L}_{\rm init} = \frac{f_{\rm inj}}{2}L_{\rm edd} + \hat{L}_B(z_0)
    \label{eq:luminositycum}.
\end{align}
This initial luminosity is shown as the gray solid line in Fig.~\ref{fig:run01_paraall}. As a final reference, we also show the AD value, as well as the evolution of the BLR and DT luminosities. 

Not surprisingly, the magnetic luminosity dominates the jet constituents. Interestingly though, it is almost constant as a function of distance. On the other hand, the proton and electron luminosities react to the different cooling strengths and the geometry of the jet, becoming constant only in the conical section. The radiative luminosity shows an interesting behavior. The initial decrease is probably related to the \g-\g\ absorption process, which is particularly strong at small $z$. In case A, a peak is visible at about $0.1\,$pc, while in case B no such peak is evident. While this points to a mild dominance of jet regions around $0.1\,$pc for the radiative output in case A, one should note that even in this case the distribution is broad, and no clear dominating emission region can be found. Nonetheless, this points towards the delicate interplay of external photon fields (their location and strength) and the acceleration of the bulk flow. In case A the peak of the radiative luminosity is located around the edge of the BLR, while in case B the edge of the BLR has already been passed.

The total luminosity is at or below the initial value at all distances. In case A, the baseline parameters result in jet power below the accretion power, while in case B the jet power is initially about a factor two higher than the accretion one. The injection fraction $f_{\rm inj}$ has been set a factor of a few below the limit of Eq.~(\ref{eq:fedd}). However, injecting at the maximally allowed limit would only marginally increased the total luminosity. Hence for the baseline simulation, \textit{ExHaLe-jet} works within bounds similar to simulations of MAD disks \citep{tchekhovskoy+11}.

%
%
\subsubsection{Neutrino emission} \label{sec:sim01neu}
\begin{figure}
\centering 
\includegraphics[width=0.48\textwidth]{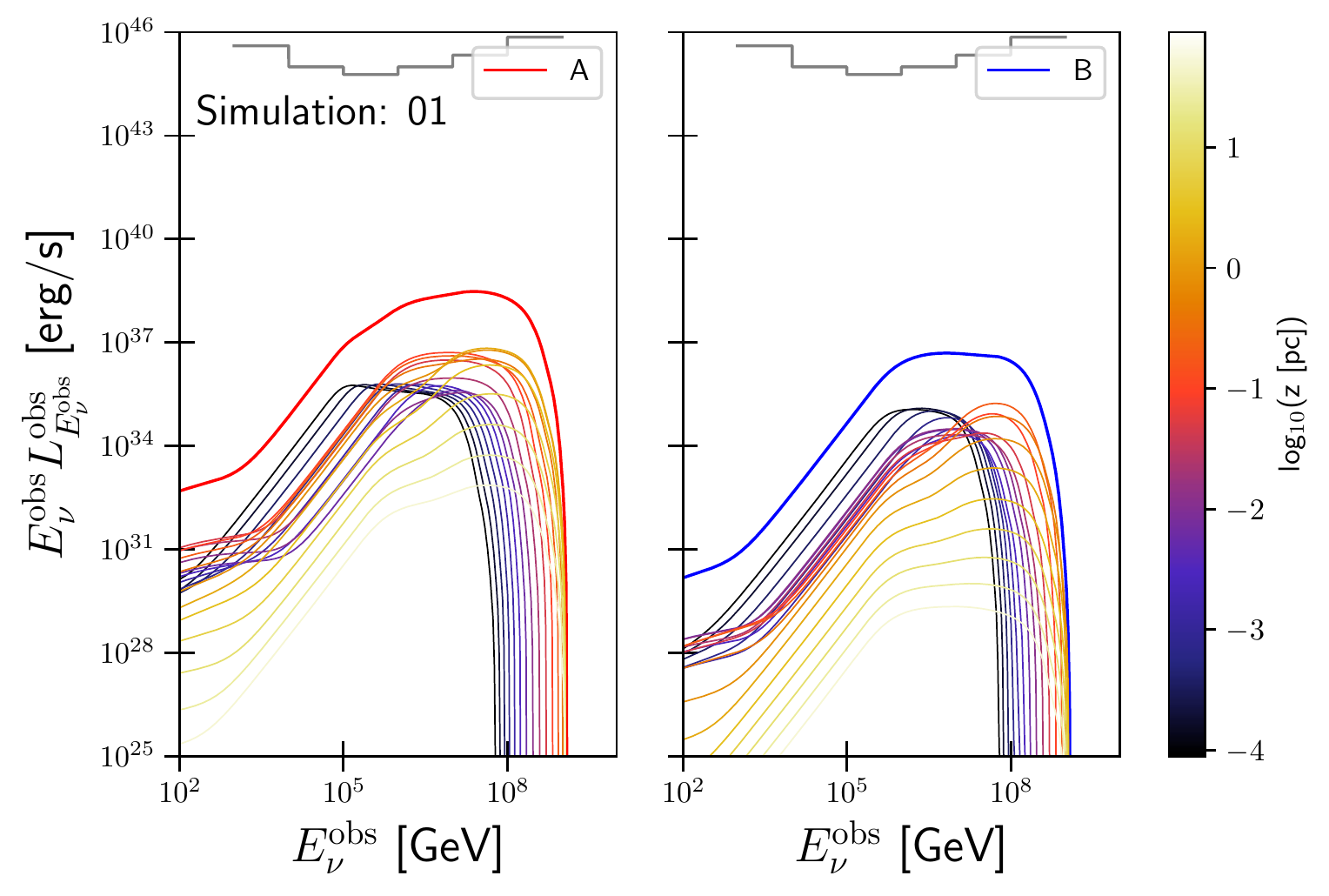}
\caption{Total muon neutrino spectra (red and blue thick solid lines) in the observer's frame and their evolution with distance $z$ (color code) as a function of energy for simulation 01 A (left) and B (right). The thin colored lines show the neutrino spectra of every tenth slice. In both panels, the gray solid line marks the expected sensitivity of IceCube-Gen2 \citep{icecubegentwo21}.
}
\label{fig:run01_neutrino}
\end{figure} 
The decay of charged pions and muons produces neutrinos. The resulting muon neutrino spectrum in the observer's frame and its evolution with distance $z$ for the baseline simulation is shown in Fig.~\ref{fig:run01_neutrino}. 
We also show the planned sensitivity of the IceCube-Gen2 detector \citep{icecubegentwo21}. With the given parameter set of our jet, no neutrino detection is expected. It is nonetheless instructive to consider the differences in the neutrino spectra of case A and B.

In case of strong external fields a higher number of neutrinos is obtained than for weak external fields. While the peaks of the distributions are attained at roughly similar energies of $\sim 10^{8}\,$GeV (observer frame), the spectral shape at lower energies is different. In the strong-disk case, the neutrinos are mostly produced on distances between $0.1$ and a few pc from the black hole. Interestingly, the spectrum produced at $\sim 0.1\,$pc is broader and peaks at lower energies than the spectrum produced beyond $1\,$pc. This again reflects the relative importance between BLR and DT photons. In the weak-disk case, the neutrinos are mostly produced within $1\,$pc from the black hole. In this case, AD and DT photons are important as indicated by the distance evolution of the neutrino spectrum.

%
%
\subsection{Parameter study} \label{sec:parastudy}
\begin{figure*}
\centering 
\includegraphics[width=0.90\textwidth]{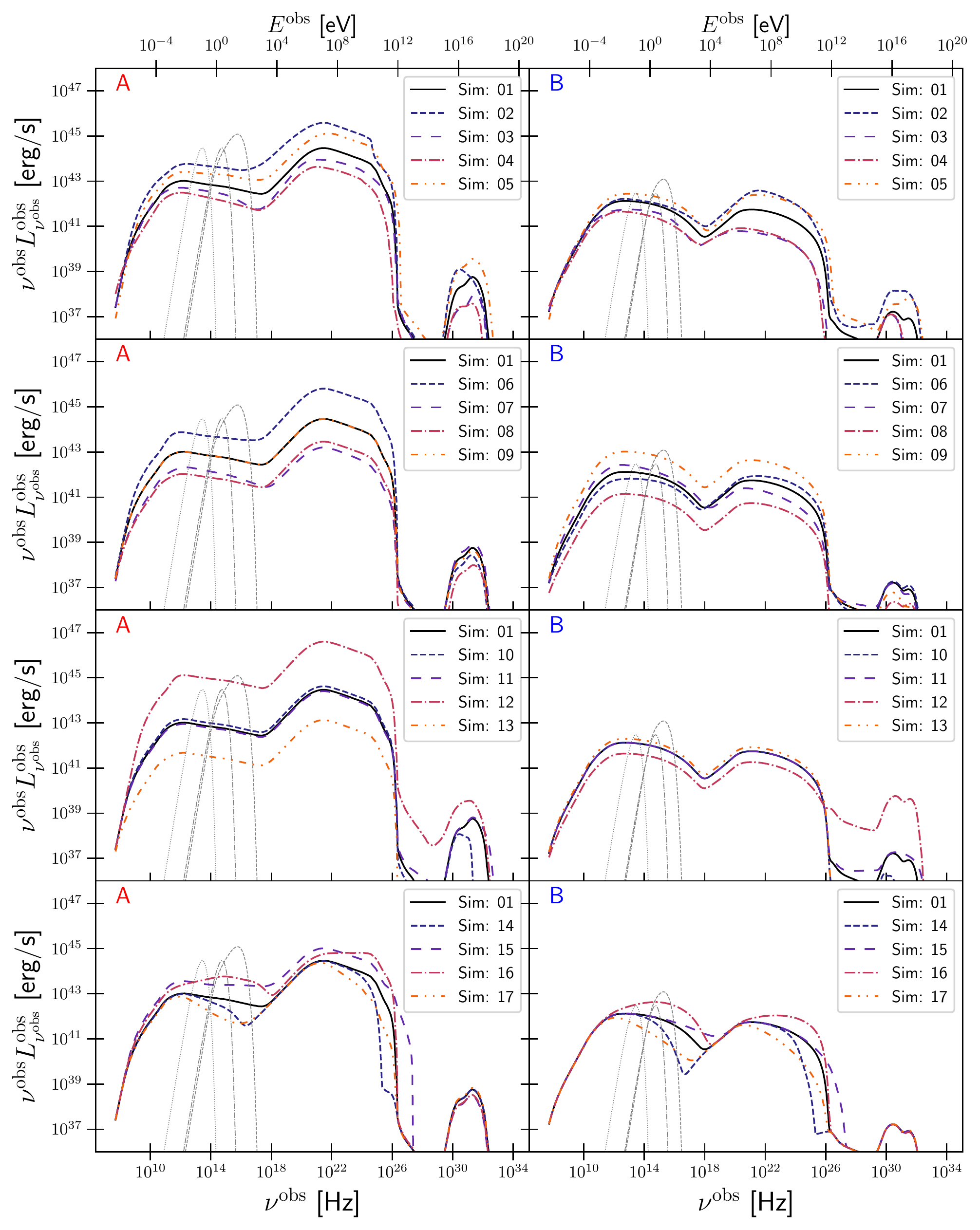}
\caption{Total photon spectra in the observer's frame for every simulation as labeled with strong-disk simulations in the left column and weak-disk simulations in the right column. Gray lines mark the AD (dashed), BLR (dash-dotted) and DT (dotted). The baseline simulation 01 (black solid) is shown in all panels for reference.
}
\label{fig:run01_17_spec}
\end{figure*} 
\begin{figure*}
\centering 
\includegraphics[width=0.90\textwidth]{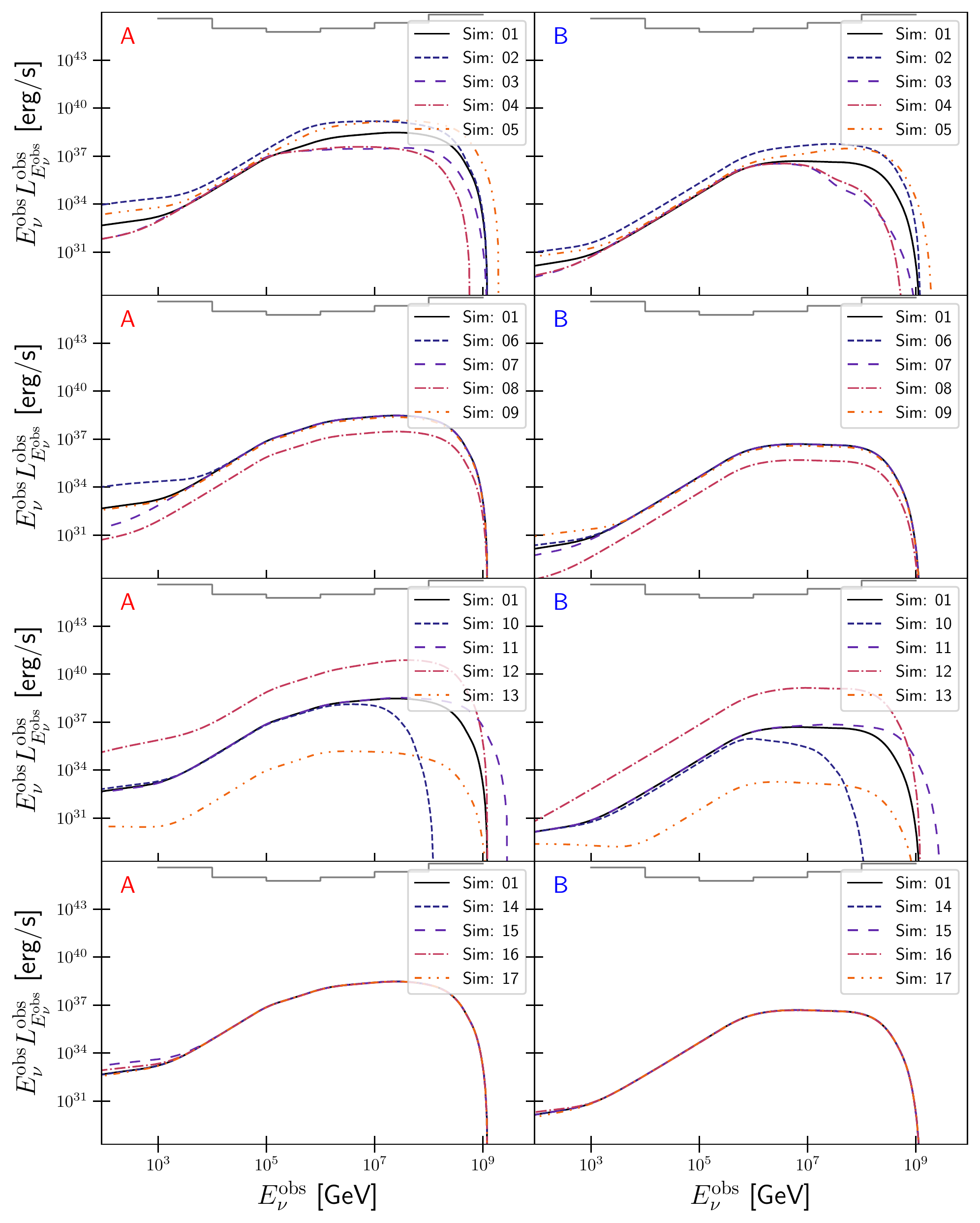}
\caption{Total muon neutrino spectra in the observer's frame for every simulation as labeled with strong-disk simulations in the left column and weak-disk simulations in the right column. The baseline simulation 01 (black solid) is shown in all panels for reference. In all panels, the gray solid line marks the expected sensitivity of IceCube-Gen2 \citep{icecubegentwo21}.
}
\label{fig:run01_17_neut}
\end{figure*} 
In this section, we compare the 16 additional simulations with the parameter variations as listed in Tab.~\ref{tab:freepara} to the baseline simulation 01. We will not go into too much detail, but merely compare the total photon spectra and total neutrino spectra. These are shown in Figs.~\ref{fig:run01_17_spec} and \ref{fig:run01_17_neut}, respectively. Table~\ref{tab:magkappa} in App.~\ref{app:figs} lists the numerical values of the magnetization $\sigma_B$ and the proton-to-electron ration $\kappa_{pe}$ at the base $z_0$, at the end of the bulk acceleration region $z_{\rm acc}$, and at the end of our jet $z_{\rm term}$. All simulation are done each for the strong-disk (A) and for the weak-disk (B) case. The magnetization is the same for cases A and B for each simulation, while for most B simulations $\kappa_{pe}$ remains at or close to the initial value. The latter implies a limited pair production due to the weak external fields.

The length of the acceleration region $z_{\rm acc}$ of the bulk flow (simulations 02 and 03) affects mainly the overall normalization, with a shorter (02) acceleration region increasing the flux, while a longer (03) acceleration region decreases it. This is true for both cases A and B, as well as for the neutrinos. The shorter acceleration region length implies that the jet reaches its maximum bulk speed deeper within the external photon fields implying a greater efficiency for IC and proton-photon processes, while a longer acceleration region has the opposite effect. 

Similar statements can be made for the variation in the maximum bulk Lorentz factor $\Gamma_{b,{\rm max}}$ of the jet flow (simulations 04 and 05). A lower $\Gamma_{b,{\rm max}}$ (04) reduces the overall normalization, while a higher one (05) increases it. A mild effect is also seen on the cut-off energy of the neutrino spectra, where a smaller $\Gamma_{b,{\rm max}}$ reduces the cut-off energy, while a higher $\Gamma_{b,{\rm max}}$ increases it. We note that we also changed the observation angles in order to ensure that $\Gamma_{b,{\rm max}}=\delta_{\rm max}$.

The variation of the magnetic field (simulations 06 and 07) results in more complicated changes. A reduction in the magnetic field (06) increases the overall flux in case A. As we keep the particle distribution fixed, the reduced synchrotron cooling results in more high-energy particles being available and, hence, producing more highly energetic radiation. In turn, more pairs are being created than in the baseline simulation as can be seen in Tab.~\ref{tab:magkappa}. The reduced external photon flux in case B implies a much reduced production of pairs. In turn, the spectral change is closer to expectation with a reduced synchrotron and a slightly increased IC flux. A higher magnetic field (07) has the opposite effect. The increased synchrotron cooling results in a weaker production of pairs compared to the baseline, which in turn means a reduced overall flux in case A. In case B, we see a higher synchrotron and a reduced IC flux as expected. The change in magnetic field only has a weak impact on pion and neutrino production with no significant change in either flux except at low neutrino energies for the here considered value of $\gamma_{p,2}$.

The reduction in the particle injection power by a factor $10$ (simulation 08) merely reduces the overall normalization by roughly an order of magnitude, as expected. On the other hand, an initial increase of the electron density by a factor $10$ (simulation 09) increases the electron-related emission by about the same factor in case B. Interestingly, there is no change to the photon spectrum in case A compared to the baseline. As indicated in Tab.~\ref{tab:magkappa}, the initial value of $\kappa_{pe}=0.1$ has almost no effect on $\kappa_{pe}$ at larger distances. In simulation 09, there is no significant change in the neutrino spectrum except at low energies.

A decrease in the maximum proton Lorentz factor $\gamma_{p,2}$ (simulation 10) has no effect on the two main peaks in the spectrum, but decreases the electron-synchrotron flux beyond $1\,$TeV. Additionally, the neutral-pion bump and the neutrino spectrum cut off at lower energies. On the other hand, increasing $\gamma_{p,2}$ (simulation 11) results in an increase in the electron-synchrotron emission beyond $1\,$TeV, as well as an increase in the neutral-pion and neutrino cut off energies. There is no difference in these effects between cases A and B.

A harder proton injection distribution (simulation 12) has severe consequences. The resulting increase in the amount of highly energetic protons enhances their interactions -- most notably by the major pair cascade (cf. Tab.~\ref{tab:magkappa}), increasing in case A the electron synchrotron and IC fluxes by three orders of magnitude compared to the baseline model. In case B, the two main bumps show a mildly reduced flux, as the harder proton spectrum implies a reduced injection normalization, Eq.~(\ref{eq:injlum00}). With the weak external fields, the pair production is much reduced compared to case A. Nonetheless, the effect of highly energetic pairs being injected is evident by the high flux beyond $1\,$TeV, which is electron-synchrotron emission (cf. Fig.~\ref{fig:run12_speccompdist}). Interestingly, the neutral pion bump exhibits higher fluxes in case B than in case A owing to a much lower degree of absorption.
The neutrino spectra increase considerably compared to the other simulations. However even in this set-up, it does not come close to the IceCube-Gen2 sensitivity. 
Softening the proton injection distribution (simulation 13) reduces the amount of high-energetic protons, therefore reducing the amount of secondaries, see Tab.~\ref{tab:magkappa}. In turn, all photon and neutrino spactra are much reduced in case A compared to the baseline simulation 01. In case B, the reduced proton-photon interactions imply much less pions and neutrinos, while the two main bumps in the SED remain almost unchanged compared to the baseline.

%
%
Decreasing the maximum electron Lorentz factor $\gamma_{e,2}$ (simulation 14) only reduces the flux at the high-energy ends of the first and second hump in the SED. Increasing $\gamma_{e,2}$ (simulation 15) has the opposite effect, while in case A even a higher normalization of the two main SED bumps is realized. With the IC emission reaching higher energies, more pairs are produced increasing the pair load of the jet (Tab.~\ref{tab:magkappa}). Reducing the electron spectral index $p_e$ (simulation 16) affects notably the first and second SED components. Interestingly, the nearly broken power-law shapes of the synchrotron peak in both cases A and B are not intuitively expected. In fact, these peaks are influenced even by slices at considerable distance from the black hole close to $z_{\rm term}$. At these distances, the magnetic field is so low that the peak frequency has shifted from the X-ray domain into the optical domain explaining why the total peak is located in that energy band (cf. Fig.~\ref{fig:run16_specdist}). The flat \g-ray peak, on the other hand, is indeed just a consequence of the chosen spectral index. An increase of $p_e$ (simulation 17) merely results in a softening of the spectra. In these simulations (14-17) only primary electron parameters have been changed. Hence, changes in the neutral-pion bump and the neutrino spectra are minor, as expected.

%
%
\section{Summary and Outlook} \label{sec:discussion}
\textit{ExHaLe-jet} is a kinetic, lepto-hadronic emission code, which models the radiation produced along the extended flow of a blazar jet. In this paper, we have introduced the code and provided a parameter study. For an efficient calculation, the jet is cut into numerous slices, wherein the particle distributions and the radiation spectra are derived. This is similar to previous purely leptonic extended jet codes \citep[e.g.,][]{pc13,zdziarski+14,lucchini+19}. The slices are connected by an assumed geometry and bulk flow profile. We also consider the presence of external photon fields, such as the AD, the BLR, the DT, and the CMB.

The crucial addition compared to the aforementioned leptonic codes is the presence of highly relativistic protons. Their interactions with ambient photons (via pion and Bethe-Heitler pair production) initiate an electromagnetic cascade (driven by \g-\g\ pair production) resulting in the accumulation of highly relativistic pairs in the jet. As these pairs are stable particles, they are carried along in the jet flow and become primary particles downstream. This has the effect that the ratio of protons to pairs decreases with distance from the black hole explaining the observed ratio in jets and lobes \citep{sikora+20}.

We have conducted a first parameter study. Within the assumed parameter range we find that the photon spectra are dominated by leptonic emission processes, namely synchrotron and IC scattering of external photon fields. Beyond a few TeV, electron-synchrotron emission of the cascade can be seen, while at ultra-high frequencies (beyond $10^{30}\,$Hz) the neutral pion bump is evident. This shows that the protons still have an effect in these set-ups even though their direct (synchrotron) emission is not visible. However, the influence of the protons depends strongly on the external fields, as the resulting effects are much more pronounced for bright external fields than for weak ones. The produced neutrinos are also not sufficient to allow individual sources to be detected by current and future neutrino instruments, such as IceCube-Gen2.


According to \cite{boccardi+21} and \cite{park+21}, jets reach their terminal velocity (or the break point from the parabolic to the conical geometry) between $10^4$ to $10^6\,R_g$. In our simulations, the jet reaches this point at $1\,$pc from the black hole corresponding to about $5\E{4}R_g$. As is shown with simulations 02 and 03, this range has significant consequences on the photon and neutrino spectra. Unfortunately, neither the reason for this range nor its relation to external entities (such as the BLR, DT, or other gas distributions) is yet known; but it could point towards important constraints on the jet and its surrounding.
We also point out that we have terminated our jet calculation at $100\,$pc, even though the emission of more distant jet regions may also be important \citep[e.g.,][]{zw16,roychowdhury+22}.

In order to improve the code, we plan several additions and amendments. First of all, except for the entrainment of the pairs produced in the cascade, the slices are almost independent of each other. Most notably, the produced radiation of one slice has no effect on other slices nor is it attenuated in downstream slices. These are crucial processes, which we are going to deal with in a subsequent paper. We will also add the production and evolution of neutrons, which may play a crucial role in the energy distribution along the jet \citep{muecke+00}. Additionally, we have neglected the light of the host galaxy. While its influence on the total spectrum through IC emission is minor \citep[cf.,][]{pottercotter13c} in most cases \citep[even though it may play a role in Centaurus~A,][]{hess20}, the host galaxy light may serve as a target for \g-ray absorption at TeV energies, which could be observable with the future Cherenkov Telescope Array \citep{zcw17}. We also plan to include a more realistic particle acceleration scenario to go beyond the current simplistic injection of the same power-law shape in each slice. Further development plans include time-dependent models to explain the observed variability \citep[as in, e.g.,][]{malzac14,potter18}, as well as radially-dependent structures in order to explain, for example, the limb-brightening seen in radio maps of several jets.

%
%
\section*{Acknowledgement}
The authors thank the anonymous referee for valuable suggestions to clarify the presentation of this manuscript.
We wish to thank Anton Dmytriiev for his invaluable help with the implementation of the Chang\&Cooper routine. We also thank Markus B\"ottcher, Matteo Cerruti, Patrick Kilian, Zakaria Meliani, and Felix Spanier for fruitful discussions.
MZ acknowledges postdoctoral financial support from LUTH, Observatoire de Paris.
AR acknowledges financial support from the Austrian Science Fund (FWF) under grant agreement number I 4144-N27.
Simulations for this paper have been performed on the TAU-cluster of the Centre for Space Research at North-West University, Potchesftroom, South Africa.

%
%
\section*{Data availability}
At this point, the code is not yet meant for public use. However, collaborations are possible on reasonable request to the corresponding author.

%
%
\bibliographystyle{mnras}
\bibliography{extendedjet}

\begin{thebibliography}{}
\makeatletter
\relax
\def\mn@urlcharsother{\let\do\@makeother \do\$\do\&\do\#\do\^\do\_\do\%\do\~}
\def\mn@doi{\begingroup\mn@urlcharsother \@ifnextchar [ {\mn@doi@}
  {\mn@doi@[]}}
\def\mn@doi@[#1]#2{\def\@tempa{#1}\ifx\@tempa\@empty \href
  {http://dx.doi.org/#2} {doi:#2}\else \href {http://dx.doi.org/#2} {#1}\fi
  \endgroup}
\def\mn@eprint#1#2{\mn@eprint@#1:#2::\@nil}
\def\mn@eprint@arXiv#1{\href {http://arxiv.org/abs/#1} {{\tt arXiv:#1}}}
\def\mn@eprint@dblp#1{\href {http://dblp.uni-trier.de/rec/bibtex/#1.xml}
  {dblp:#1}}
\def\mn@eprint@#1:#2:#3:#4\@nil{\def\@tempa {#1}\def\@tempb {#2}\def\@tempc
  {#3}\ifx \@tempc \@empty \let \@tempc \@tempb \let \@tempb \@tempa \fi \ifx
  \@tempb \@empty \def\@tempb {arXiv}\fi \@ifundefined
  {mn@eprint@\@tempb}{\@tempb:\@tempc}{\expandafter \expandafter \csname
  mn@eprint@\@tempb\endcsname \expandafter{\@tempc}}}

\bibitem[\protect\citeauthoryear{{Ackermann} et~al.,}{{Ackermann}
  et~al.}{2016}]{fermi16}
{Ackermann} M.,  et~al., 2016, \mn@doi [\apjl] {10.3847/2041-8205/824/2/L20},
  \href {https://ui.adsabs.harvard.edu/abs/2016ApJ...824L..20A} {824, L20}

\bibitem[\protect\citeauthoryear{{Aharonian}, {Atoian}  \&
  {Nagapetian}}{{Aharonian} et~al.}{1983}]{aan83}
{Aharonian} F.~A.,  {Atoian} A.~M.,   {Nagapetian} A.~M.,  1983, Astrofizika,
  \href {https://ui.adsabs.harvard.edu/abs/1983Afz....19..323A} {19, 323}

\bibitem[\protect\citeauthoryear{{Aharonian} et~al.,}{{Aharonian}
  et~al.}{2007}]{hess07pks}
{Aharonian} F.,  et~al., 2007, \mn@doi [\apjl] {10.1086/520635}, \href
  {https://ui.adsabs.harvard.edu/abs/2007ApJ...664L..71A} {664, L71}

\bibitem[\protect\citeauthoryear{{Ahnen} et~al.,}{{Ahnen}
  et~al.}{2017}]{magic17}
{Ahnen} M.~L.,  et~al., 2017, \mn@doi [\aap] {10.1051/0004-6361/201629960},
  \href {https://ui.adsabs.harvard.edu/abs/2017A&A...603A..29A} {603, A29}

\bibitem[\protect\citeauthoryear{{Barr}, {Gaisser}, {Lipari}  \&
  {Tilav}}{{Barr} et~al.}{1988}]{barr+88}
{Barr} S.,  {Gaisser} T.~K.,  {Lipari} P.,   {Tilav} S.,  1988, \mn@doi
  [Physics Letters B] {10.1016/0370-2693(88)90468-6}, \href
  {https://ui.adsabs.harvard.edu/abs/1988PhLB..214..147B} {214, 147}

\bibitem[\protect\citeauthoryear{{Blandford} \& {Znajek}}{{Blandford} \&
  {Znajek}}{1977}]{blandfordznajek77}
{Blandford} R.~D.,  {Znajek} R.~L.,  1977, \mn@doi [\mnras]
  {10.1093/mnras/179.3.433}, \href
  {https://ui.adsabs.harvard.edu/abs/1977MNRAS.179..433B} {179, 433}

\bibitem[\protect\citeauthoryear{{Blumenthal}}{{Blumenthal}}{1970}]{blumenthal70}
{Blumenthal} G.~R.,  1970, \mn@doi [\prd] {10.1103/PhysRevD.1.1596}, \href
  {https://ui.adsabs.harvard.edu/abs/1970PhRvD...1.1596B} {1, 1596}

\bibitem[\protect\citeauthoryear{{Boccardi} et~al.,}{{Boccardi}
  et~al.}{2021}]{boccardi+21}
{Boccardi} B.,  et~al., 2021, \mn@doi [\aap] {10.1051/0004-6361/202039612},
  \href {https://ui.adsabs.harvard.edu/abs/2021A&A...647A..67B} {647, A67}

\bibitem[\protect\citeauthoryear{{Boettcher}, {Harris}  \&
  {Krawczynski}}{{Boettcher} et~al.}{2012}]{bhk12}
{Boettcher} M.,  {Harris} D.~E.,   {Krawczynski} H.,  2012, {Relativistic Jets
  from Active Galactic Nuclei}.
Wiley-VCH

\bibitem[\protect\citeauthoryear{{B{\"o}ttcher} \& {Els}}{{B{\"o}ttcher} \&
  {Els}}{2016}]{be16}
{B{\"o}ttcher} M.,  {Els} P.,  2016, \mn@doi [\apj]
  {10.3847/0004-637X/821/2/102}, \href
  {https://ui.adsabs.harvard.edu/abs/2016ApJ...821..102B} {821, 102}

\bibitem[\protect\citeauthoryear{{B{\"o}ttcher}, {Mause}  \&
  {Schlickeiser}}{{B{\"o}ttcher} et~al.}{1997}]{boettcher+97}
{B{\"o}ttcher} M.,  {Mause} H.,   {Schlickeiser} R.,  1997, \aap, 324, 395

\bibitem[\protect\citeauthoryear{{Casadio} et~al.,}{{Casadio}
  et~al.}{2021}]{casadio+21}
{Casadio} C.,  et~al., 2021, \mn@doi [\aap] {10.1051/0004-6361/202039616},
  \href {https://ui.adsabs.harvard.edu/abs/2021A&A...649A.153C} {649, A153}

\bibitem[\protect\citeauthoryear{{Celotti} \& {Fabian}}{{Celotti} \&
  {Fabian}}{1993}]{celottifabian93}
{Celotti} A.,  {Fabian} A.~C.,  1993, \mn@doi [\mnras]
  {10.1093/mnras/264.1.228}, \href
  {https://ui.adsabs.harvard.edu/abs/1993MNRAS.264..228C} {264, 228}

\bibitem[\protect\citeauthoryear{{Celotti} \& {Ghisellini}}{{Celotti} \&
  {Ghisellini}}{2008}]{celottighiesellini08}
{Celotti} A.,  {Ghisellini} G.,  2008, \mn@doi [\mnras]
  {10.1111/j.1365-2966.2007.12758.x}, \href
  {https://ui.adsabs.harvard.edu/abs/2008MNRAS.385..283C} {385, 283}

\bibitem[\protect\citeauthoryear{{Cerruti}}{{Cerruti}}{2020}]{cerruti20}
{Cerruti} M.,  2020, \mn@doi [Galaxies] {10.3390/galaxies8040072}, \href
  {https://ui.adsabs.harvard.edu/abs/2020Galax...8...72C} {8, 72}

\bibitem[\protect\citeauthoryear{{Cerruti}, {Zech}, {Boisson}, {Emery}, {Inoue}
   \& {Lenain}}{{Cerruti} et~al.}{2019}]{cerruti+19}
{Cerruti} M.,  {Zech} A.,  {Boisson} C.,  {Emery} G.,  {Inoue} S.,   {Lenain}
  J.~P.,  2019, \mn@doi [\mnras] {10.1093/mnrasl/sly210}, \href
  {https://ui.adsabs.harvard.edu/abs/2019MNRAS.483L..12C} {483, L12}

\bibitem[\protect\citeauthoryear{{Cerruti}, {Zech}, {Boisson}, {Emery}, {Inoue}
   \& {Lenain}}{{Cerruti} et~al.}{2021}]{czbea21}
{Cerruti} M.,  {Zech} A.,  {Boisson} C.,  {Emery} G.,  {Inoue} S.,   {Lenain}
  J.~P.,  2021, \mn@doi [\mnras] {10.1093/mnrasl/slaa188}, \href
  {https://ui.adsabs.harvard.edu/abs/2021MNRAS.502L..21C} {502, L21}

\bibitem[\protect\citeauthoryear{{Chang} \& {Cooper}}{{Chang} \&
  {Cooper}}{1970}]{changcooper70}
{Chang} J.,  {Cooper} G.,  1970, \mn@doi [Journal of Computational Physics]
  {10.1016/0021-9991(70)90001-X}, 6, 1

\bibitem[\protect\citeauthoryear{{Chatterjee} et~al.,}{{Chatterjee}
  et~al.}{2020}]{chatterjee+20}
{Chatterjee} K.,  et~al., 2020, \mn@doi [\mnras] {10.1093/mnras/staa2718},
  \href {https://ui.adsabs.harvard.edu/abs/2020MNRAS.499..362C} {499, 362}

\bibitem[\protect\citeauthoryear{{Chen}, {Pohl}  \& {B{\"o}ttcher}}{{Chen}
  et~al.}{2015}]{chen+15}
{Chen} X.,  {Pohl} M.,   {B{\"o}ttcher} M.,  2015, \mn@doi [\mnras]
  {10.1093/mnras/stu2438}, \href
  {https://ui.adsabs.harvard.edu/abs/2015MNRAS.447..530C} {447, 530}

\bibitem[\protect\citeauthoryear{{Chiaberge} \& {Ghisellini}}{{Chiaberge} \&
  {Ghisellini}}{1999}]{chiabergeghisellini99}
{Chiaberge} M.,  {Ghisellini} G.,  1999, \mn@doi [\mnras]
  {10.1046/j.1365-8711.1999.02538.x}, \href
  {https://ui.adsabs.harvard.edu/abs/1999MNRAS.306..551C} {306, 551}

\bibitem[\protect\citeauthoryear{{Chodorowski}, {Zdziarski}  \&
  {Sikora}}{{Chodorowski} et~al.}{1992}]{chodorowski+92}
{Chodorowski} M.~J.,  {Zdziarski} A.~A.,   {Sikora} M.,  1992, \mn@doi [\apj]
  {10.1086/171984}, \href
  {https://ui.adsabs.harvard.edu/abs/1992ApJ...400..181C} {400, 181}

\bibitem[\protect\citeauthoryear{{Dermer} \& {Menon}}{{Dermer} \&
  {Menon}}{2009}]{dm09}
{Dermer} C.~D.,  {Menon} G.,  2009, {High Energy Radiation from Black Holes:
  Gamma Rays, Cosmic Rays, and Neutrinos}.
Princeton Univerisity Press

\bibitem[\protect\citeauthoryear{{Diltz} \& {B{\"o}ttcher}}{{Diltz} \&
  {B{\"o}ttcher}}{2014}]{db14}
{Diltz} C.,  {B{\"o}ttcher} M.,  2014, \mn@doi [Journal of High Energy
  Astrophysics] {10.1016/j.jheap.2014.04.001}, \href
  {https://ui.adsabs.harvard.edu/abs/2014JHEAp...1...63D} {1, 63}

\bibitem[\protect\citeauthoryear{{Dmytriiev}, {Sol}  \& {Zech}}{{Dmytriiev}
  et~al.}{2021}]{dmytriiev+21}
{Dmytriiev} A.,  {Sol} H.,   {Zech} A.,  2021, \mn@doi [\mnras]
  {10.1093/mnras/stab1445}, \href
  {https://ui.adsabs.harvard.edu/abs/2021MNRAS.505.2712D} {505, 2712}

\bibitem[\protect\citeauthoryear{{Dong}, {Zhang}  \& {Giannios}}{{Dong}
  et~al.}{2020}]{dong+20}
{Dong} L.,  {Zhang} H.,   {Giannios} D.,  2020, \mn@doi [\mnras]
  {10.1093/mnras/staa773}, \href
  {https://ui.adsabs.harvard.edu/abs/2020MNRAS.494.1817D} {494, 1817}

\bibitem[\protect\citeauthoryear{{Fichet de Clairfontaine}, {Meliani}, {Zech}
  \& {Hervet}}{{Fichet de Clairfontaine} et~al.}{2021}]{fichet+21}
{Fichet de Clairfontaine} G.,  {Meliani} Z.,  {Zech} A.,   {Hervet} O.,  2021,
  \mn@doi [\aap] {10.1051/0004-6361/202039654}, \href
  {https://ui.adsabs.harvard.edu/abs/2021A&A...647A..77F} {647, A77}

\bibitem[\protect\citeauthoryear{{Gaisser}}{{Gaisser}}{1990}]{gaisser90}
{Gaisser} T.~K.,  1990, {Cosmic rays and particle physics.}.
Cambridge University Press

\bibitem[\protect\citeauthoryear{{Gao}, {Fedynitch}, {Winter}  \& {Pohl}}{{Gao}
  et~al.}{2019}]{gao+19}
{Gao} S.,  {Fedynitch} A.,  {Winter} W.,   {Pohl} M.,  2019, \mn@doi [Nature
  Astronomy] {10.1038/s41550-018-0610-1}, \href
  {https://ui.adsabs.harvard.edu/abs/2019NatAs...3...88G} {3, 88}

\bibitem[\protect\citeauthoryear{{Ghisellini} \& {Tavecchio}}{{Ghisellini} \&
  {Tavecchio}}{2008}]{gt08}
{Ghisellini} G.,  {Tavecchio} F.,  2008, \mn@doi [\mnras]
  {10.1111/j.1365-2966.2008.13360.x}, \href
  {https://ui.adsabs.harvard.edu/abs/2008MNRAS.387.1669G} {387, 1669}

\bibitem[\protect\citeauthoryear{{Ghisellini} \& {Tavecchio}}{{Ghisellini} \&
  {Tavecchio}}{2010}]{ghisellinitavecchio10}
{Ghisellini} G.,  {Tavecchio} F.,  2010, \mn@doi [\mnras]
  {10.1111/j.1745-3933.2010.00952.x}, \href
  {https://ui.adsabs.harvard.edu/abs/2010MNRAS.409L..79G} {409, L79}

\bibitem[\protect\citeauthoryear{{Ghisellini}, {Celotti}, {George}  \&
  {Fabian}}{{Ghisellini} et~al.}{1992}]{ghisellini+92}
{Ghisellini} G.,  {Celotti} A.,  {George} I.~M.,   {Fabian} A.~C.,  1992,
  \mn@doi [\mnras] {10.1093/mnras/258.4.776}, \href
  {https://ui.adsabs.harvard.edu/abs/1992MNRAS.258..776G} {258, 776}

\bibitem[\protect\citeauthoryear{{H.E.S.S. Collaboration} et~al.,}{{H.E.S.S.
  Collaboration} et~al.}{2019}]{hess19}
{H.E.S.S. Collaboration} et~al., 2019, \mn@doi [\aap]
  {10.1051/0004-6361/201935704}, \href
  {https://ui.adsabs.harvard.edu/abs/2019A&A...627A.159H} {627, A159}

\bibitem[\protect\citeauthoryear{{H.E.S.S. Collaboration} et~al.,}{{H.E.S.S.
  Collaboration} et~al.}{2020}]{hess20}
{H.E.S.S. Collaboration} et~al., 2020, \mn@doi [\nat]
  {10.1038/s41586-020-2354-1}, \href
  {https://ui.adsabs.harvard.edu/abs/2020Natur.582..356H} {582, 356}

\bibitem[\protect\citeauthoryear{{H.E.S.S. Collaboration} et~al.,}{{H.E.S.S.
  Collaboration} et~al.}{2021}]{hess+21}
{H.E.S.S. Collaboration} et~al., 2021, \mn@doi [\aap]
  {10.1051/0004-6361/202038949}, \href
  {https://ui.adsabs.harvard.edu/abs/2021A&A...648A..23H} {648, A23}

\bibitem[\protect\citeauthoryear{{Harris} \& {Krawczynski}}{{Harris} \&
  {Krawczynski}}{2006}]{hk06}
{Harris} D.~E.,  {Krawczynski} H.,  2006, \mn@doi [\araa]
  {10.1146/annurev.astro.44.051905.092446}, \href
  {https://ui.adsabs.harvard.edu/abs/2006ARA&A..44..463H} {44, 463}

\bibitem[\protect\citeauthoryear{{Hayashida} et~al.,}{{Hayashida}
  et~al.}{2012}]{hayashida+12}
{Hayashida} M.,  et~al., 2012, \mn@doi [\apj] {10.1088/0004-637X/754/2/114},
  \href {https://ui.adsabs.harvard.edu/abs/2012ApJ...754..114H} {754, 114}

\bibitem[\protect\citeauthoryear{{Hervet}, {Boisson}  \& {Sol}}{{Hervet}
  et~al.}{2015}]{hervet+15}
{Hervet} O.,  {Boisson} C.,   {Sol} H.,  2015, \mn@doi [\aap]
  {10.1051/0004-6361/201425330}, \href
  {https://ui.adsabs.harvard.edu/abs/2015A&A...578A..69H} {578, A69}

\bibitem[\protect\citeauthoryear{{Hovatta} et~al.,}{{Hovatta}
  et~al.}{2021}]{hovatta+21}
{Hovatta} T.,  et~al., 2021, \mn@doi [\aap] {10.1051/0004-6361/202039481},
  \href {https://ui.adsabs.harvard.edu/abs/2021A&A...650A..83H} {650, A83}

\bibitem[\protect\citeauthoryear{{H{\"u}mmer}, {R{\"u}ger}, {Spanier}  \&
  {Winter}}{{H{\"u}mmer} et~al.}{2010}]{huemmer+10}
{H{\"u}mmer} S.,  {R{\"u}ger} M.,  {Spanier} F.,   {Winter} W.,  2010, \mn@doi
  [\apj] {10.1088/0004-637X/721/1/630}, \href
  {https://ui.adsabs.harvard.edu/abs/2010ApJ...721..630H} {721, 630}

\bibitem[\protect\citeauthoryear{{IceCube Collaboration} et~al.,}{{IceCube
  Collaboration} et~al.}{2018}]{icecube+18}
{IceCube Collaboration} et~al., 2018, \mn@doi [Science]
  {10.1126/science.aat1378}, \href
  {https://ui.adsabs.harvard.edu/abs/2018Sci...361.1378I} {361, eaat1378}

\bibitem[\protect\citeauthoryear{{Kantzas} et~al.,}{{Kantzas}
  et~al.}{2021}]{kantzas+21}
{Kantzas} D.,  et~al., 2021, \mn@doi [\mnras] {10.1093/mnras/staa3349}, \href
  {https://ui.adsabs.harvard.edu/abs/2021MNRAS.500.2112K} {500, 2112}

\bibitem[\protect\citeauthoryear{{Kelner} \& {Aharonian}}{{Kelner} \&
  {Aharonian}}{2008}]{ka08}
{Kelner} S.~R.,  {Aharonian} F.~A.,  2008, \mn@doi [\prd]
  {10.1103/PhysRevD.78.034013}, \href
  {https://ui.adsabs.harvard.edu/abs/2008PhRvD..78c4013K} {78, 034013}

\bibitem[\protect\citeauthoryear{{K\"onigl}}{{K\"onigl}}{1980}]{koenigl80}
{K\"onigl} A.,  1980, Phys Fluids, 23, 1083

\bibitem[\protect\citeauthoryear{{Lucchini}, {Markoff}, {Crumley}, {Krau{\ss}}
  \& {Connors}}{{Lucchini} et~al.}{2019}]{lucchini+19}
{Lucchini} M.,  {Markoff} S.,  {Crumley} P.,  {Krau{\ss}} F.,   {Connors}
  R.~M.~T.,  2019, \mn@doi [\mnras] {10.1093/mnras/sty2929}, \href
  {https://ui.adsabs.harvard.edu/abs/2019MNRAS.482.4798L} {482, 4798}

\bibitem[\protect\citeauthoryear{{Malzac}}{{Malzac}}{2014}]{malzac14}
{Malzac} J.,  2014, \mn@doi [\mnras] {10.1093/mnras/stu1144}, \href
  {https://ui.adsabs.harvard.edu/abs/2014MNRAS.443..299M} {443, 299}

\bibitem[\protect\citeauthoryear{{M{\"u}cke}, {Engel}, {Rachen}, {Protheroe}
  \& {Stanev}}{{M{\"u}cke} et~al.}{2000}]{muecke+00}
{M{\"u}cke} A.,  {Engel} R.,  {Rachen} J.~P.,  {Protheroe} R.~J.,   {Stanev}
  T.,  2000, \mn@doi [Computer Physics Communications]
  {10.1016/S0010-4655(99)00446-4}, \href
  {https://ui.adsabs.harvard.edu/abs/2000CoPhC.124..290M} {124, 290}

\bibitem[\protect\citeauthoryear{{M{\"u}cke}, {Protheroe}, {Engel}, {Rachen}
  \& {Stanev}}{{M{\"u}cke} et~al.}{2003}]{muecke+03}
{M{\"u}cke} A.,  {Protheroe} R.~J.,  {Engel} R.,  {Rachen} J.~P.,   {Stanev}
  T.,  2003, \mn@doi [Astroparticle Physics] {10.1016/S0927-6505(02)00185-8},
  \href {https://ui.adsabs.harvard.edu/abs/2003APh....18..593M} {18, 593}

\bibitem[\protect\citeauthoryear{{Netzer}}{{Netzer}}{2015}]{netzer15}
{Netzer} H.,  2015, \mn@doi [\araa] {10.1146/annurev-astro-082214-122302},
  \href {https://ui.adsabs.harvard.edu/abs/2015ARA&A..53..365N} {53, 365}

\bibitem[\protect\citeauthoryear{{Park} \& {Petrosian}}{{Park} \&
  {Petrosian}}{1996}]{parkpetrosian96}
{Park} B.~T.,  {Petrosian} V.,  1996, \mn@doi [\apjs] {10.1086/192278}, \href
  {https://ui.adsabs.harvard.edu/abs/1996ApJS..103..255P} {103, 255}

\bibitem[\protect\citeauthoryear{{Park}, {Hada}, {Nakamura}, {Asada}, {Zhao}
  \& {Kino}}{{Park} et~al.}{2021}]{park+21}
{Park} J.,  {Hada} K.,  {Nakamura} M.,  {Asada} K.,  {Zhao} G.,   {Kino} M.,
  2021, \mn@doi [\apj] {10.3847/1538-4357/abd6ee}, \href
  {https://ui.adsabs.harvard.edu/abs/2021ApJ...909...76P} {909, 76}

\bibitem[\protect\citeauthoryear{{Pepe}, {Vila}  \& {Romero}}{{Pepe}
  et~al.}{2015}]{pepe+15}
{Pepe} C.,  {Vila} G.~S.,   {Romero} G.~E.,  2015, \mn@doi [\aap]
  {10.1051/0004-6361/201527156}, \href
  {https://ui.adsabs.harvard.edu/abs/2015A&A...584A..95P} {584, A95}

\bibitem[\protect\citeauthoryear{{Potter}}{{Potter}}{2018}]{potter18}
{Potter} W.~J.,  2018, \mn@doi [\mnras] {10.1093/mnras/stx2371}, \href
  {https://ui.adsabs.harvard.edu/abs/2018MNRAS.473.4107P} {473, 4107}

\bibitem[\protect\citeauthoryear{{Potter} \& {Cotter}}{{Potter} \&
  {Cotter}}{2013a}]{pc13}
{Potter} W.~J.,  {Cotter} G.,  2013a, \mn@doi [\mnras] {10.1093/mnras/sts407},
  \href {https://ui.adsabs.harvard.edu/abs/2013MNRAS.429.1189P} {429, 1189}

\bibitem[\protect\citeauthoryear{{Potter} \& {Cotter}}{{Potter} \&
  {Cotter}}{2013b}]{pottercotter13c}
{Potter} W.~J.,  {Cotter} G.,  2013b, \mn@doi [\mnras] {10.1093/mnras/stt1569},
  \href {https://ui.adsabs.harvard.edu/abs/2013MNRAS.436..304P} {436, 304}

\bibitem[\protect\citeauthoryear{{Press}, {Flannery}, {Teukolsky}  \&
  {Vetterling}}{{Press} et~al.}{1989}]{press+89}
{Press} W.~H.,  {Flannery} B.~P.,  {Teukolsky} S.~A.,   {Vetterling} W.~T.,
  1989, {Numerical recipes in Pascal. The art of scientific computing}.
Cambridge University Press

\bibitem[\protect\citeauthoryear{{Pushkarev}, {Kovalev}, {Lister}  \&
  {Savolainen}}{{Pushkarev} et~al.}{2009}]{pushkarev+09}
{Pushkarev} A.~B.,  {Kovalev} Y.~Y.,  {Lister} M.~L.,   {Savolainen} T.,  2009,
  \mn@doi [\aap] {10.1051/0004-6361/200913422}, \href
  {https://ui.adsabs.harvard.edu/abs/2009A&A...507L..33P} {507, L33}

\bibitem[\protect\citeauthoryear{{Pushkarev}, {Kovalev}, {Lister}  \&
  {Savolainen}}{{Pushkarev} et~al.}{2017}]{pushkarev+17}
{Pushkarev} A.~B.,  {Kovalev} Y.~Y.,  {Lister} M.~L.,   {Savolainen} T.,  2017,
  \mn@doi [\mnras] {10.1093/mnras/stx854}, \href
  {https://ui.adsabs.harvard.edu/abs/2017MNRAS.468.4992P} {468, 4992}

\bibitem[\protect\citeauthoryear{{Reimer}, {B{\"o}ttcher}  \& {Buson}}{{Reimer}
  et~al.}{2019}]{reimer+19}
{Reimer} A.,  {B{\"o}ttcher} M.,   {Buson} S.,  2019, \mn@doi [\apj]
  {10.3847/1538-4357/ab2bff}, \href
  {https://ui.adsabs.harvard.edu/abs/2019ApJ...881...46R} {881, 46}

\bibitem[\protect\citeauthoryear{{Roychowdhury}, {Meyer}, {Georganopoulos},
  {Breiding}  \& {Petropoulou}}{{Roychowdhury} et~al.}{2021}]{roychowdhury+22}
{Roychowdhury} A.,  {Meyer} E.~T.,  {Georganopoulos} M.,  {Breiding} P.,
  {Petropoulou} M.,  2021, arXiv e-prints, \href
  {https://ui.adsabs.harvard.edu/abs/2021arXiv211012016R} {p. arXiv:2110.12016}

\bibitem[\protect\citeauthoryear{{Schlickeiser}}{{Schlickeiser}}{2002}]{rsch02}
{Schlickeiser} R.,  2002, {Cosmic Ray Astrophysics}.
Springer Verlag

\bibitem[\protect\citeauthoryear{{Shakura} \& {Sunyaev}}{{Shakura} \&
  {Sunyaev}}{1973}]{ss73}
{Shakura} N.~I.,  {Sunyaev} R.~A.,  1973, \aap, \href
  {https://ui.adsabs.harvard.edu/abs/1973A&A....24..337S} {500, 33}

\bibitem[\protect\citeauthoryear{{Sikora} \& {Madejski}}{{Sikora} \&
  {Madejski}}{2000}]{sikoramadejski00}
{Sikora} M.,  {Madejski} G.,  2000, \mn@doi [\apj] {10.1086/308756}, \href
  {https://ui.adsabs.harvard.edu/abs/2000ApJ...534..109S} {534, 109}

\bibitem[\protect\citeauthoryear{{Sikora}, {Nalewajko}  \& {Madejski}}{{Sikora}
  et~al.}{2020}]{sikora+20}
{Sikora} M.,  {Nalewajko} K.,   {Madejski} G.~M.,  2020, \mn@doi [\mnras]
  {10.1093/mnras/staa3128}, \href
  {https://ui.adsabs.harvard.edu/abs/2020MNRAS.499.3749S} {499, 3749}

\bibitem[\protect\citeauthoryear{{Tchekhovskoy}, {Narayan}  \&
  {McKinney}}{{Tchekhovskoy} et~al.}{2011}]{tchekhovskoy+11}
{Tchekhovskoy} A.,  {Narayan} R.,   {McKinney} J.~C.,  2011, \mn@doi [\mnras]
  {10.1111/j.1745-3933.2011.01147.x}, \href
  {https://ui.adsabs.harvard.edu/abs/2011MNRAS.418L..79T} {418, L79}

\bibitem[\protect\citeauthoryear{{The IceCube-Gen2 Collaboration} et~al.,}{{The
  IceCube-Gen2 Collaboration} et~al.}{2020}]{icecubegentwo21}
{The IceCube-Gen2 Collaboration} et~al., 2020, arXiv e-prints, \href
  {https://ui.adsabs.harvard.edu/abs/2020arXiv200804323T} {p. arXiv:2008.04323}

\bibitem[\protect\citeauthoryear{{Vercellone} et~al.,}{{Vercellone}
  et~al.}{2011}]{vercellone+11}
{Vercellone} S.,  et~al., 2011, \mn@doi [\apjl] {10.1088/2041-8205/736/2/L38},
  \href {https://ui.adsabs.harvard.edu/abs/2011ApJ...736L..38V} {736, L38}

\bibitem[\protect\citeauthoryear{{Vila}, {Romero}  \& {Casco}}{{Vila}
  et~al.}{2012}]{vila+12}
{Vila} G.~S.,  {Romero} G.~E.,   {Casco} N.~A.,  2012, \mn@doi [\aap]
  {10.1051/0004-6361/201118106}, \href
  {https://ui.adsabs.harvard.edu/abs/2012A&A...538A..97V} {538, A97}

\bibitem[\protect\citeauthoryear{{Weidinger} \& {Spanier}}{{Weidinger} \&
  {Spanier}}{2015}]{weidingerspanier15}
{Weidinger} M.,  {Spanier} F.,  2015, \mn@doi [\aap]
  {10.1051/0004-6361/201424159}, \href
  {https://ui.adsabs.harvard.edu/abs/2015A&A...573A...7W} {573, A7}

\bibitem[\protect\citeauthoryear{{Zacharias} \& {Wagner}}{{Zacharias} \&
  {Wagner}}{2016}]{zw16}
{Zacharias} M.,  {Wagner} S.~J.,  2016, \mn@doi [\aap]
  {10.1051/0004-6361/201526698}, \href
  {https://ui.adsabs.harvard.edu/abs/2016A&A...588A.110Z} {588, A110}

\bibitem[\protect\citeauthoryear{{Zacharias}, {Chen}  \& {Wagner}}{{Zacharias}
  et~al.}{2017a}]{zcw17}
{Zacharias} M.,  {Chen} X.,   {Wagner} S.~J.,  2017a, \mn@doi [\mnras]
  {10.1093/mnras/stw3032}, \href
  {https://ui.adsabs.harvard.edu/abs/2017MNRAS.465.3767Z} {465, 3767}

\bibitem[\protect\citeauthoryear{{Zacharias}, {B{\"o}ttcher}, {Jankowsky},
  {Lenain}, {Wagner}  \& {Wierzcholska}}{{Zacharias}
  et~al.}{2017b}]{zacharias+17}
{Zacharias} M.,  {B{\"o}ttcher} M.,  {Jankowsky} F.,  {Lenain} J.~P.,  {Wagner}
  S.~J.,   {Wierzcholska} A.,  2017b, \mn@doi [\apj]
  {10.3847/1538-4357/aa9bee}, \href
  {https://ui.adsabs.harvard.edu/abs/2017ApJ...851...72Z} {851, 72}

\bibitem[\protect\citeauthoryear{{Zdziarski} \& {Bottcher}}{{Zdziarski} \&
  {Bottcher}}{2015}]{zb15}
{Zdziarski} A.~A.,  {Bottcher} M.,  2015, \mn@doi [\mnras]
  {10.1093/mnrasl/slv039}, \href
  {https://ui.adsabs.harvard.edu/abs/2015MNRAS.450L..21Z} {450, L21}

\bibitem[\protect\citeauthoryear{{Zdziarski}, {Stawarz}, {Pjanka}  \&
  {Sikora}}{{Zdziarski} et~al.}{2014}]{zdziarski+14}
{Zdziarski} A.~A.,  {Stawarz} {\L}.,  {Pjanka} P.,   {Sikora} M.,  2014,
  \mn@doi [\mnras] {10.1093/mnras/stu420}, \href
  {https://ui.adsabs.harvard.edu/abs/2014MNRAS.440.2238Z} {440, 2238}

\bibitem[\protect\citeauthoryear{{Zdziarski}, {Sikora}, {Pjanka}  \&
  {Tchekhovskoy}}{{Zdziarski} et~al.}{2015}]{Zdziarski+15}
{Zdziarski} A.~A.,  {Sikora} M.,  {Pjanka} P.,   {Tchekhovskoy} A.,  2015,
  \mn@doi [\mnras] {10.1093/mnras/stv986}, \href
  {https://ui.adsabs.harvard.edu/abs/2015MNRAS.451..927Z} {451, 927}

\makeatother
\end{thebibliography}
%
%
\appendix
\section{Derivation of the injection normalization and the injection fraction} \label{app:partmag}
In order to derive Eq.~(\ref{eq:injlum00}) we remind ourselves that the injection luminosity equals the integrated injection rate (which is total energy density per unit time) times the volume in which the particles are injected. Assuming a power-law distribution of the injected particles of species $i$ (protons or electrons) between a lower ($\gamma_{i,1}$) and upper ($\gamma_{i,2}$) cut-off, we find

\begin{align}
    Q_i(\gamma) = q_{i} \gamma^{-p_{i}} \SF{\gamma}{\gamma_{i,1}}{\gamma_{i,2}}
    \label{eq:app1},
\end{align}
where $\SF{x}{a}{b}$ is unity for $a\leq x\leq b$ and zero otherwise. The total density of particles (considering escape) then becomes

\begin{align}
    n_i = q_{i}t_{\rm esc}\intl_{\gamma_{i,1}}^{\gamma_{i,2}}\gamma^{-p_{i}}\td{\gamma} = q_{i}t_{\rm esc}\mathcal{I}_{i0}
    \label{eq:app2},
\end{align}
while the total energy density can be written as

\begin{align}
    u_i = q_{i}t_{\rm esc}m_ic^2\intl_{\gamma_{i,1}}^{\gamma_{i,2}}\gamma^{1-p_{i}}\td{\gamma} = q_{i}t_{\rm esc}m_ic^2\mathcal{I}_{i1}
    \label{eq:app3}.
\end{align}
In both equations we have employed the definition of the integral, Eq.~(\ref{eq:injectintegrals}).

Relating the proton and electron densities with $\kappa_{pe}$ provides us with:

\begin{align}
    n_e &= q_{e}t_{\rm esc}\mathcal{I}_{e0} \stackrel{!}{=} \frac{n_p}{\kappa_{pe}} = \frac{q_{p}t_{\rm esc}\mathcal{I}_{p0}}{\kappa_{pe}} \label{eq:app4} \\
    \Leftrightarrow q_{e} &= \frac{q_{p}\mathcal{I}_{p0}}{\kappa_{pe}\mathcal{I}_{e0}} \label{eq:app5}.
\end{align}
The total energy density then becomes

\begin{align}
    u_p+u_e &= q_{p}t_{\rm esc}m_pc^2\mathcal{I}_{p1} + q_{e}t_{\rm esc}m_ec^2\mathcal{I}_{e1} \nonumber \\
    &= q_{p}t_{\rm esc} \left[ m_pc^2\mathcal{I}_{p1} + \frac{m_ec^2}{\kappa_{pe}\mathcal{I}_{e0}}\mathcal{I}_{p0}\mathcal{I}_{e1} \right]
    \label{eq:app6}.
\end{align}
Dividing Eq.~(\ref{eq:app6}) by $t_{\rm esc}$ to obtain the energy density per unit time, setting $q_{p}\equiv q$, and multiplying with the volume of the base slice (at $z_0$) provides Eq.~\ref{eq:injlum00}.


The total power $L_{\rm tot}$ injected into the base of the jet (until the end of this section we only consider quantities at $z=z_0$) from the accretion process is distributed in particles and magnetic field:

\begin{align}
    L_{\rm tot} = L_{\rm inj} + \pi \eta_R^2z_0^2 c u_B = \frac{f_{\rm tot}L_{\rm edd}}{2\Gamma_{b,0}^2}
    \label{eq:app7}.
\end{align}
Here, $f_{\rm tot}$ is the fraction of the Eddington power injected into the jet in the form of both particles and magnetic field. As $L_{\rm inj}>0$, we immediately obtain a lower limit on $f_{\rm tot}$:

\begin{align}
    f_{\rm tot}> f_{\rm min} := \frac{c\Gamma_{b,0}^2\eta_R^2z_0^2B^2}{4L_{\rm edd}}
    \label{eq:app_fmin}.
\end{align}

From the condition on the ``unperturbed'' flow, $\sigma_B>(\Gamma_{b,{\rm max}}/\Gamma_{b,0})-1$, and the definition of the magnetization, Eq.~(\ref{eq:magnetization}), we obtain

\begin{align}
    2u_B &> \left( \frac{\Gamma_{b,{\rm max}}}{\Gamma_{b,0}}-1 \right) \left[ \eta_{\rm ad} u+\rho c^2 \right] \nonumber \\
    &= \left( \frac{\Gamma_{b,{\rm max}}}{\Gamma_{b,0}}-1 \right) qt_{\rm esc} \nonumber \\
    &\quad\times \left[ \eta_{\rm ad} \left( m_pc^2\mathcal{I}_{p1} + \frac{m_ec^2}{\kappa_{pe}\mathcal{I}_{e0}}\mathcal{I}_{p0}\mathcal{I}_{e1} \right) \right. \nonumber \\
    &\quad\left. + m_pc^2\mathcal{I}_{p0} + \frac{m_ec^2}{\kappa_{pe}\mathcal{I}_{e0}}\mathcal{I}_{p0}\mathcal{I}_{e0} \right]
    \label{eq:app8}.
\end{align}
Inserting $q$ and $t_{\rm esc}$ as given in Secs.~\ref{sec:large} and \ref{sec:slices}, and defining

\begin{align}
    \zeta &:= \frac{\eta_{\rm ad} \left( m_pc^2\mathcal{I}_{p1} + \frac{m_ec^2}{\kappa_{pe}\mathcal{I}_{e0}}\mathcal{I}_{p0}\mathcal{I}_{e1} \right) + m_pc^2\mathcal{I}_{p0} + \frac{m_ec^2}{\kappa_{pe}}\mathcal{I}_{p0}}{m_pc^2\mathcal{I}_{p1} + \frac{m_ec^2}{\kappa_{pe}\mathcal{I}_{e0}}\mathcal{I}_{p0}\mathcal{I}_{e1}} \nonumber \\
    &= \eta_{\rm ad} + \frac{\mathcal{I}_{p0} \left( 1+\frac{m_e}{m_p\kappa_{pe}} \right)}{\mathcal{I}_{p1} + \frac{m_e}{m_p\kappa_{pe}\mathcal{I}_{e0}}\mathcal{I}_{p0}\mathcal{I}_{e1}}
    \label{eq:app_zeta},
\end{align}
we obtain

\begin{align}
    u_B \left[ 2 + \left( \frac{\Gamma_{b,{\rm max}}}{\Gamma_{b,0}}-1 \right)\eta_{\rm esc}\zeta \right] \nonumber \\
    > \left( \frac{\Gamma_{b,{\rm max}}}{\Gamma_{b,0}}-1 \right) \frac{f_{\rm tot}L_{\rm edd}\eta_{\rm esc}\zeta}{2c\Gamma_{b,0}^2\pi \eta_R^2z_0^2}
    \label{eq:app9}.
\end{align}
Solving for $f_{\rm tot}$, we obtain an upper limit:

\begin{align}
    f_{\rm tot} < f_{\rm max} := f_{\rm min} \left[ 1 + \frac{2}{\left( \frac{\Gamma_{b,{\rm max}}}{\Gamma_{b,0}}-1 \right)\eta_{\rm esc}\zeta} \right]
    \label{eq:app_fmax}.
\end{align}
As $f_{\rm tot}$ contains the contribution of both the particle and the magnetic power, we can set the particle fraction $f_{\rm inj}$ to

\begin{align}
    f_{inj} := f_{\rm tot} - f_{\rm min}
    \label{eq:app_finj},
\end{align}
which immediately transforms Eq.~(\ref{eq:app7}) into Eq.~(\ref{eq:injlum01}), and the upper limit, Eq.~(\ref{eq:app_fmax}), into Eq.~(\ref{eq:fedd}).

It is instructive to discuss the implication of the limited range of the injection power. The lower limit is derived from the simple demand that the particle content is larger than $0$, while the upper limit is a consequence of the Bernoulli equation in combination with the ``unperturbed flow'' approximation. Both limits are separated by the second summand in Eq.~(\ref{eq:app_fmax}), which we shall refer to as $f_2$ in this paragraph. It basically defines the amount of particle power that can be loaded into the jet. Recalling that $\eta_{\rm esc}>1$ and $\zeta>4/3$, $f_2$ is determined by the ratio of the maximum and initial bulk Lorentz factor. For the examples used in our parameter study, $f_2\ll 1$ (on the order of $10^{-3}$). Hence, the jet can only be loaded with a limited supply of particle power in order to be able to fully accelerate to $\Gamma_{b,{\rm max}}$ -- that is, satisfying the Bernoulli equation. Only for weakly accelerating jets ($\Gamma_{b,{\rm max}}\lesssim 5\Gamma_{b,0}/2$), $f_2$ approaches unity. For non-accelerating jets ($\Gamma_{b,{\rm max}}\rightarrow\Gamma_{b,0}$), $f_2$ approaches infinity, and the jet can be loaded with any particle power.

%
%
\section{Solving the Fokker-Planck equation} \label{app:solver}
The Fokker-Planck equation, Eq.~(\ref{eq:fpgen}), is numerically evaluated using the solver developed by \cite{changcooper70} with significant additions by \cite{parkpetrosian96} and \cite{chiabergeghisellini99}. We provide a brief overview \citep[see also][]{dmytriiev+21} here.

Equation~(\ref{eq:fpgen}) is discretized on two grids; one for the momentum $\chi$, and one for the time $t$. Designating grid points by $\chi_j$ and $t_k$, Eq.~(\ref{eq:fpgen}) can be written in the form

\begin{align}
	V1_j n_{j-1}^{k+1} + V2_j n_{j}^{k+1} + V3_j n_{j-1}^{k+1} = n_j^{k} + Q_{j}^{k}\Delta t
	\label{eq:fpnum1},
\end{align}
where $\Delta t$ is the integration time step. As we are only interested in equilibrium solutions, we set $\Delta t=10t_{\rm esc}$ ensuring the determination of the equilibrium in only a few time steps, while still running stably. The coefficients in Eq.~(\ref{eq:fpnum1}) are

\begin{align}
	V1_j &= - \frac{\Delta t}{\Delta\chi_j}\frac{C^k(\chi_{j-1/2})}{\Delta\chi_{j-1/2}}W_{j-1/2}^{-} \label{eq:fpnumV1} \\
	V2_j &= 1+\frac{\Delta t}{t_{\rm esc}}+\frac{\Delta t}{\gamma_j t\p_{\rm decay}} + \frac{\Delta t}{\Delta\chi_j}\left[ \frac{C^k(\chi_{j-1/2})}{\Delta\chi_{j-1/2}}W_{j-1/2}^{+} \right. \nonumber \\ 
	&\quad \left. + \frac{C^k(\chi_{j+1/2})}{\Delta\chi_{j+1/2}}W_{j+1/2}^{-} \right] \label{eq:fpnumV2} \\
	V3_j &= - \frac{\Delta t}{\Delta\chi_j}\frac{C^k(\chi_{j+1/2})}{\Delta\chi_{j+1/2}}W_{j+1/2}^{+} \label{eq:fpnumV3},
\end{align}
with

\begin{align}
	\Delta\chi_j &= \chi_{j+1/2} - \chi_{j-1/2} \label{eq:fpnuma} \\
	\Delta\chi_{j\pm1/2} &= \chi_{j\pm1/2+1/2} - \chi_{j\pm1/2-1/2} \label{eq:fpnumb} \\
	W_{j\pm1/2}^{\pm} &= \frac{w_{j\pm1/2} \exp{\left( \pm w_{j\pm1/2}/2 \right)}}{2\sinh{\left( w_{j\pm1/2}/2 \right)}} \label{eq:fbnumc} \\
	w_{j\pm1/2} &= \frac{B^k(\chi_{j\pm1/2})}{C^k(\chi_{j\pm1/2})}\Delta\chi_{j\pm1/2} \label{eq:fpnumd}.
\end{align}
The functions

\begin{align}
	B(\chi,t) &= |\dot{\chi}(\chi,t)| - \left[ \frac{1}{t_{\rm acc}} + \frac{2}{(2+a)t_{\rm acc}} \right] \chi \label{eq:fpnumB} \\
	C(\chi,t) &= \frac{\chi^2}{(2+a)t_{\rm acc}} \label{eq:fpnumC}
\end{align}
are evaluated at the momentum grid midpoints $\chi_{j\pm1/2}$. We note that for large absolute values of $w_{j\pm1/2}$, the functions $W_{j\pm1/2}^{\pm}$ are well approximated by $W\approx w$ or zero -- depending on the case.

Equation~(\ref{eq:fpnum1}) represents a tri-diagonal matrix, which can be solved using the steps provided in \cite{press+89}. The solution is the particle distribution $n_i(\chi)$ of a given particle species. In each time step, this routine is employed for all radiating particle species (except neutral pions). The equilibrium solution is accepted, if the total densities $n_p$ and $n_e$ of protons and electrons do not change by more than $1\times 10^{-4}$ relative to the previous two time steps. This condition ensures a stable result.

\section{Injection of secondary electron-positron pairs} \label{app:sec}
\begin{figure*}
\centering 
\includegraphics[width=0.90\textwidth]{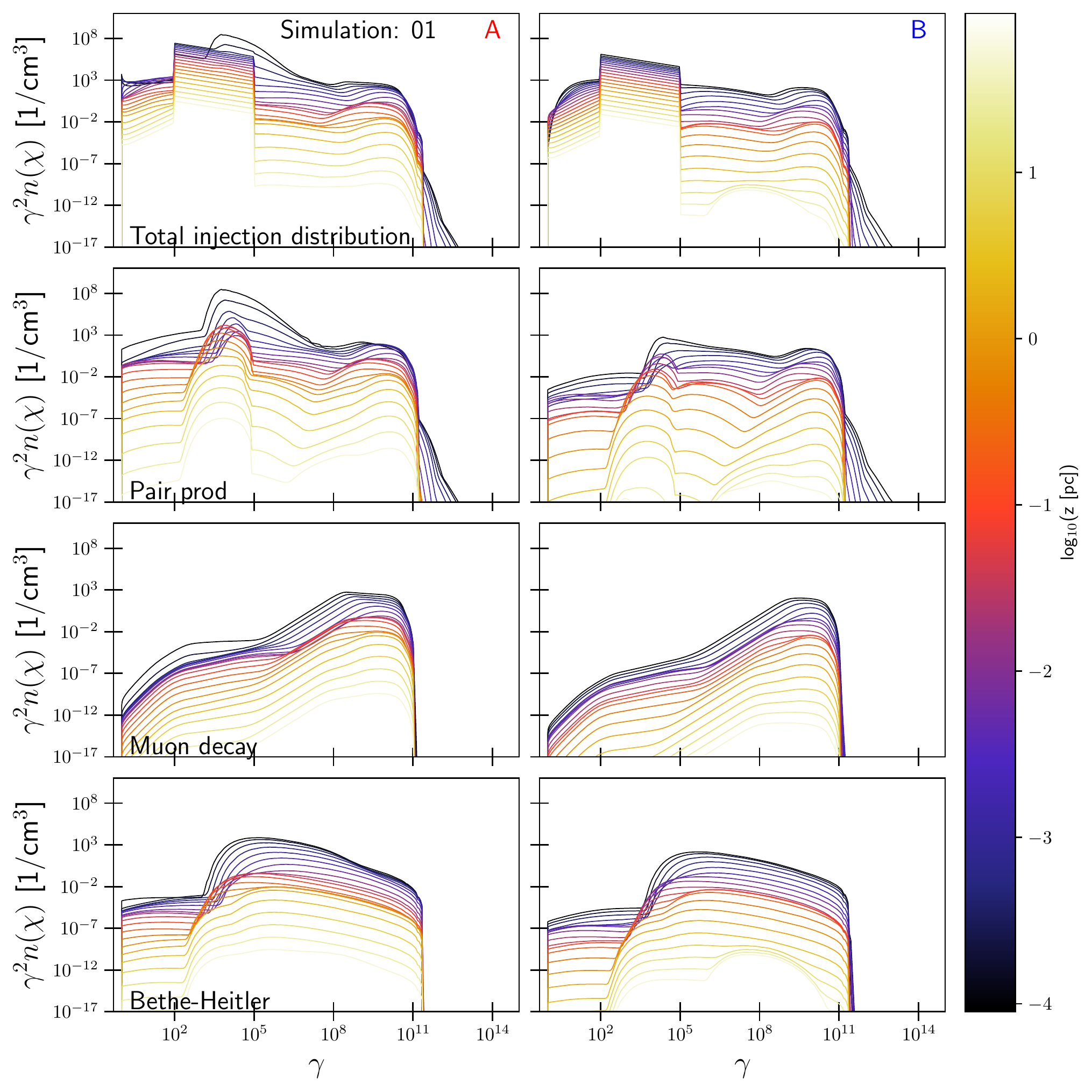}
\caption{Injection electron distribution function (top row) and secondary injection distribution functions as a function of Lorentz factor $\gamma$ and distance $z$ (color code) as labeled for simulation 01 A (left) and B (right). The lines show the spectra for every tenth slice, and are given in the comoving frame of the slice. 
}
\label{fig:run01_partseco}
\end{figure*}
In order to further discuss the influence of the secondary particles on the electron distribution function, we show in Fig.~\ref{fig:run01_partseco} the electron injection distribution and the individual injection distributions of the secondary production processes of the baseline simulation. These are the distributions before solving Eq.~(\ref{eq:fpgen}). 

The primary injection between Lorentz factors $10^2$ and $10^5$ is visible in the top row of Fig.~\ref{fig:run01_partseco}. At small Lorentz factors below $10^3$, \g-\g\ pair production dominates the secondary injection. Given that in terms of total number of particles, this energy regime provides most particles, we can deduce that overall most secondaries are injected through \g-\g\ pair production. The conditions close to the base of the jet must be very favorable for \g-\g\ pair production, as most secondary electrons are injected there. The jet is slow and the produced \g\ rays interact predominantly with AD photons. As the jet accelerates and leaves the disk behind, the number of \g-\g-pair-produces secondaries drops quickly. 

In the secondary injection spectra, Bethe-Heitler pair production becomes important in the Lorentz factor interval $10^3$ to $10^7$, as it is at least comparable to \g-\g\ pair production or may even be dominating. At higher Lorentz factors between $10^7$ and $10^{11}$ the injection from decaying muons becomes important and even the dominant process depending on distance $z$. Beyond Lorentz factors of $10^{11}$, only a few secondaries are injected from \g-\g\ pair production. While there are differences between case A and B, they are only minor.

\section{Additional figures and table} \label{app:figs}
\begin{figure*}
\centering 
\includegraphics[width=0.86\hsize]{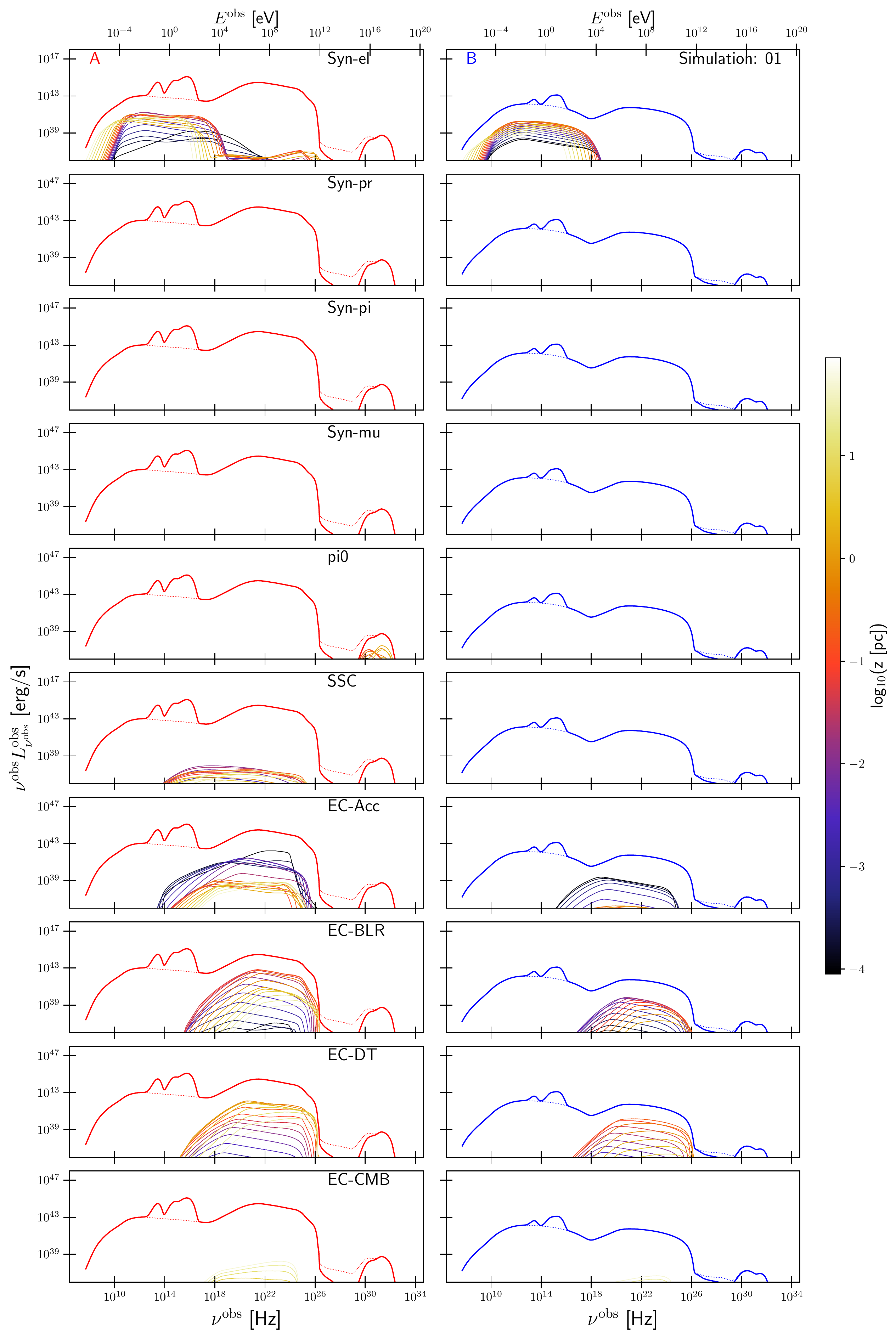}
\caption{Same as Fig.~\ref{fig:run01_specdist} but showing all individual radiative components.
}
\label{fig:run01_speccompdist}
\end{figure*}

\begin{figure*}
\centering 
\includegraphics[width=0.86\hsize]{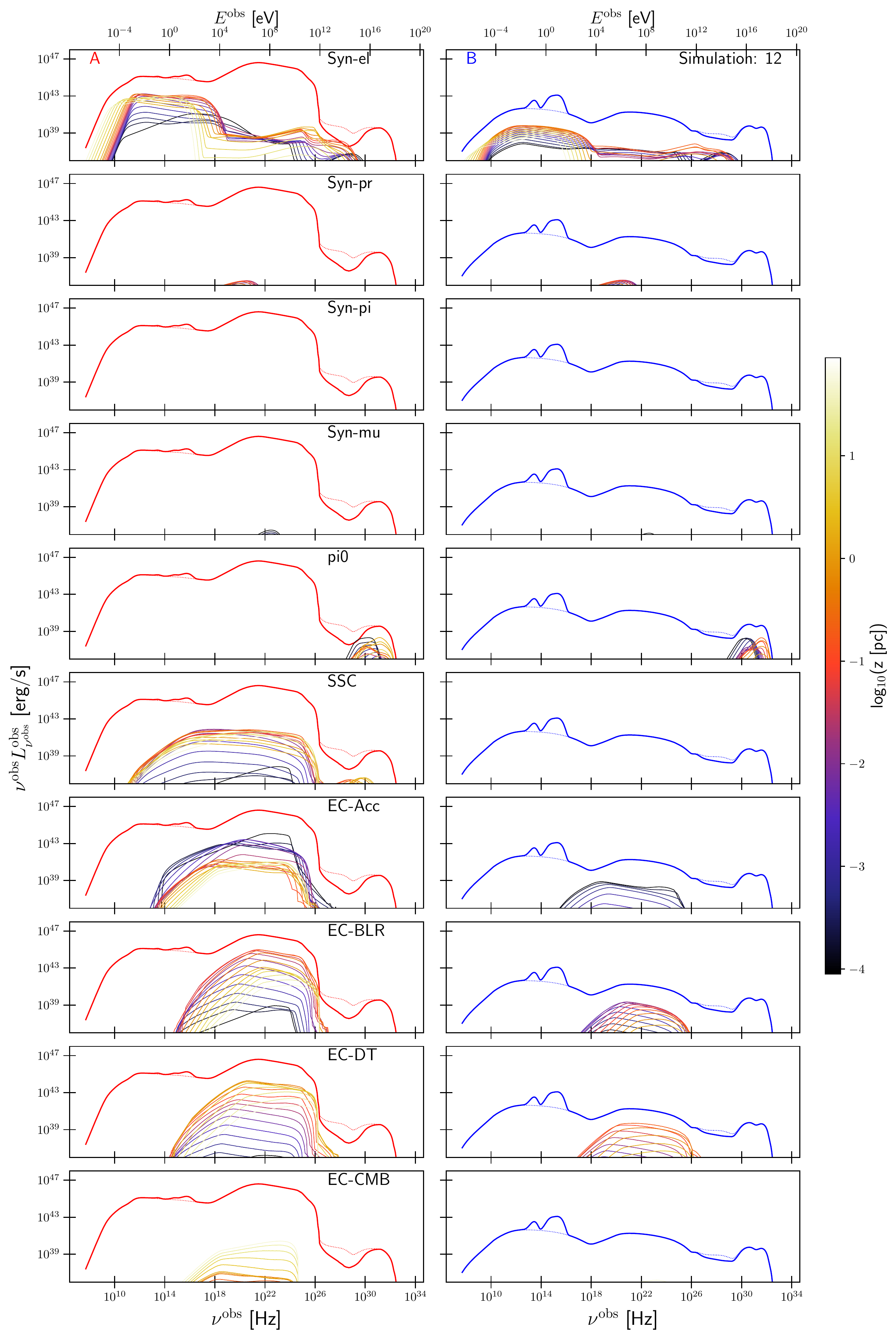}
\caption{Same as Fig.~\ref{fig:run01_speccompdist} but for simulation 12.
}
\label{fig:run12_speccompdist}
\end{figure*}

\begin{figure*}
\centering 
\includegraphics[width=0.80\textwidth]{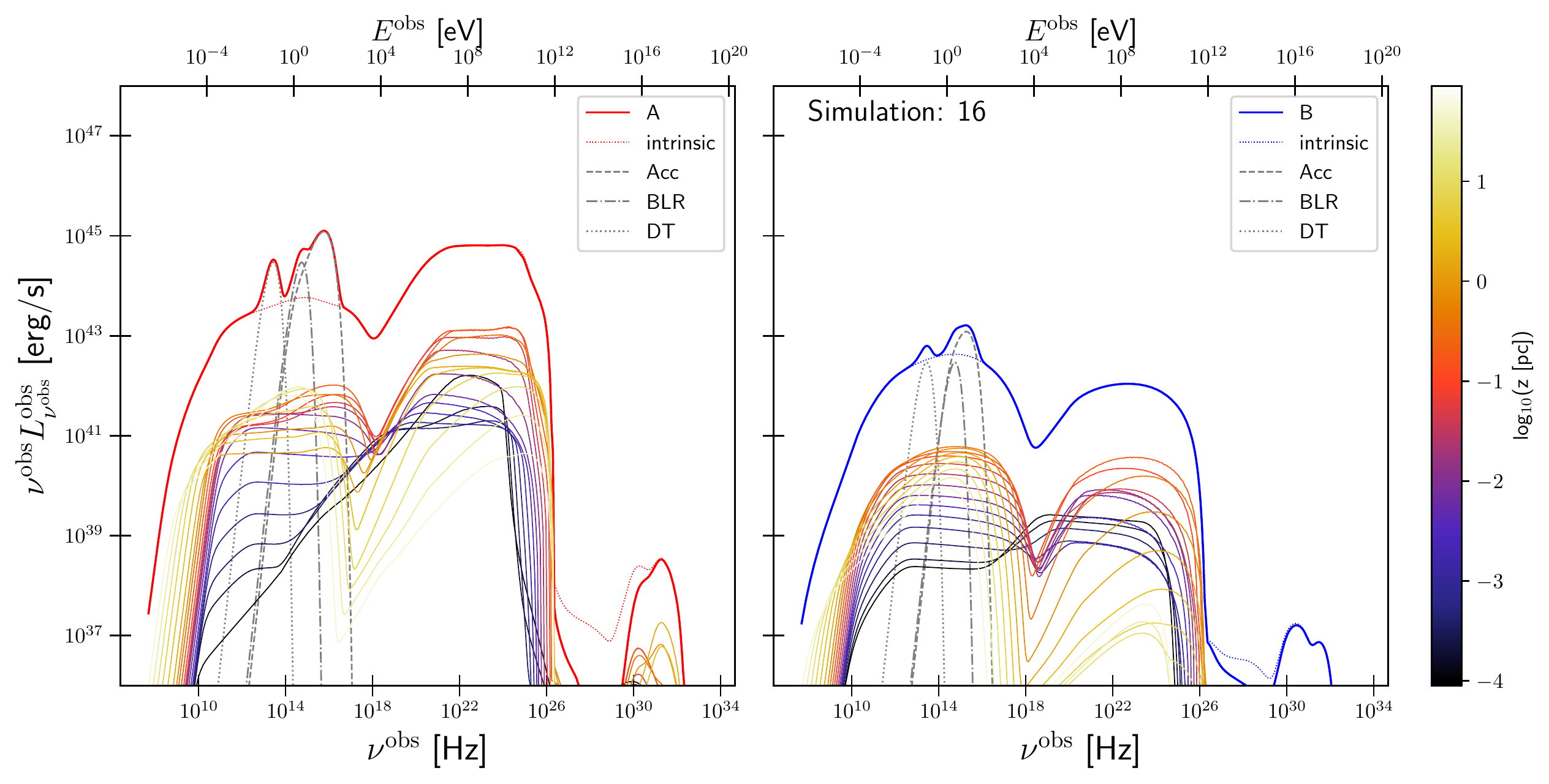}
\caption{Same as Fig.~\ref{fig:run01_specdist}, but for simulation 16.
}
\label{fig:run16_specdist}
\end{figure*}

In this section, we show three additional figures displaying the individual spectral contributions for simulations 01 (Fig.~\ref{fig:run01_speccompdist}) and 12 (Fig.~\ref{fig:run12_speccompdist}), as well as the evolution of the total spectrum of simulation 16 (Fig.~\ref{fig:run16_specdist}). Table~\ref{tab:magkappa} lists the values of $\kappa_{pe}$ and $\sigma_B$ for three distances along the jet.
%
%
%
%
%
\begin{table*}
\caption{Magnetization $\sigma_B$ and proton-to-electron ration $\kappa_{pe}$ at three locations in the jet for the various simulations. These values are calculated for the steady-state solution in the respective slices, which is why especially $\kappa_{pe}(z_0)$ can differ from the injection value given in Tab.~\ref{tab:freepara}.
}
\begin{tabular}{c|ccc|ccc}
Sim & $\sigma_B(z_0)$ & $\sigma_B(z_{\rm acc})$ & $\sigma_B(z_{\rm term})$ & $\kappa_{pe}(z_0)$ & $\kappa_{pe}(z_{\rm acc})$ & $\kappa_{pe}(z_{\rm term})$ \\
\hline
01\,A & $1.69\E{2}$ & $9.24\E{0}$ & $6.44\E{0}$ & $1.33\E{-1}$ & $2.65\E{-2}$ & $2.65\E{-2}$ \\
01\,B & $1.69\E{2}$ & $1.03\E{1}$ & $1.05\E{1}$ & 1.0 & 1.0 & 1.0 \\
02\,A & $1.72\E{2}$ & $7.29\E{0}$ & $1.61\E{0}$ & $1.02\E{-1}$ & $3.12\E{-3}$ & $2.97\E{-3}$ \\
02\,B & $1.72\E{2}$ & $8.84\E{0}$ & $1.09\E{1}$ & 1.0 & 1.0 & 1.0 \\
03\,A & $1.68\E{2}$ & $8.04\E{0}$ & $7.66\E{0}$ & $1.37\E{-1}$ & $4.61\E{-2}$ & $4.61\E{-2}$ \\
03\,B & $1.68\E{2}$ & $1.05\E{1}$ & $1.04\E{1}$ & 1.0 & 1.0 & 1.0 \\
04\,A & $1.68\E{2}$ & $2.03\E{1}$ & $1.75\E{1}$ & $1.60\E{-1}$ & $5.49\E{-2}$ & $5.49\E{-2}$ \\
04\,B & $1.68\E{2}$ & $2.25\E{1}$ & $2.26\E{1}$ & 1.0 & 1.0 & 1.0 \\
05\,A & $1.70\E{2}$ & $4.78\E{0}$ & $2.35\E{0}$ & $1.11\E{-1}$ & $1.21\E{-2}$ & $1.21\E{-2}$ \\
05\,B & $1.70\E{2}$ & $5.44\E{0}$ & $5.64\E{0}$ & 1.0 & 1.0 & 1.0 \\
06\,A & $5.85\E{1}$ & $7.04\E{-1}$ & $1.79\E{-1}$ & $7.07\E{-3}$ & $1.24\E{-3}$ & $1.24\E{-3}$ \\
06\,B & $6.07\E{1}$ & $2.62\E{0}$ & $2.67\E{0}$ & 1.0 & 1.0 & 1.0 \\
07\,A & $6.77\E{2}$ & $4.65\E{1}$ & $4.61\E{1}$ & $8.23\E{-1}$ & $4.11\E{-1}$ & $4.11\E{-1}$ \\
07\,B & $6.76\E{2}$ & $4.63\E{1}$ & $4.72\E{1}$ & 1.0 & 1.0 & 1.0 \\
08\,A & $1.69\E{3}$ & $1.06\E{2}$ & $7.42\E{1}$ & $1.33\E{-1}$ & $2.68\E{-2}$ & $2.68\E{-2}$ \\
08\,B & $1.69\E{3}$ & $1.18\E{2}$ & $1.21\E{2}$ & 1.0 & 1.0 & 1.0 \\
09\,A & $2.13\E{2}$ & $1.14\E{1}$ & $7.40\E{0}$ & $6.03\E{-2}$ & $2.10\E{-2}$ & $2.09\E{-2}$ \\
09\,B & $2.08\E{2}$ & $1.15\E{1}$ & $1.16\E{1}$ & 0.1 & 0.1 & 0.1 \\
10\,A & $1.69\E{2}$ & $8.82\E{0}$ & $5.51\E{0}$ & $9.03\E{-2}$ & $1.87\E{-2}$ & $1.87\E{-2}$ \\
10\,B & $1.69\E{2}$ & $1.03\E{1}$ & $1.05\E{1}$ & 1.0 & 1.0 & 1.0 \\
11\,A & $1.69\E{2}$ & $9.39\E{0}$ & $6.79\E{0}$ & $1.50\E{-1}$ & $3.06\E{-2}$ & $3.06\E{-2}$ \\
11\,B & $1.69\E{2}$ & $1.03\E{1}$ & $1.05\E{1}$ & 1.0 & 1.0 & 1.0 \\
12\,A & $1.71\E{2}$ & $5.17\E{-1}$ & $1.07\E{-1}$ & $5.90\E{-4}$ & $5.70\E{-5}$ & $5.69\E{-5}$ \\
12\,B & $1.89\E{2}$ & $1.44\E{1}$ & $1.49\E{1}$ & 0.99 & 0.99 & 0.99 \\
13\,A & $1.58\E{2}$ & $8.50\E{0}$ & $8.45\E{0}$ & $9.85\E{-1}$ & $8.60\E{-1}$ & $8.59\E{-1}$ \\
13\,B & $1.57\E{2}$ & $8.38\E{0}$ & $8.48\E{0}$ & 1.0 & 1.0 & 1.0 \\
14\,A & $1.69\E{2}$ & $9.31\E{0}$ & $6.80\E{0}$ & $1.33\E{-1}$ & $2.89\E{-2}$ & $2.89\E{-2}$ \\
14\,B & $1.68\E{2}$ & $1.03\E{1}$ & $1.05\E{1}$ & 1.0 & 1.0 & 1.0 \\
15\,A & $1.69\E{2}$ & $7.14\E{0}$ & $3.16\E{0}$ & $1.26\E{-1}$ & $7.45\E{-3}$ & $7.43\E{-3}$ \\
15\,B & $1.69\E{2}$ & $1.03\E{1}$ & $1.05\E{1}$ & 1.0 & 0.99 & 0.99 \\
16\,A & $1.73\E{2}$ & $8.36\E{0}$ & $3.21\E{0}$ & $1.32\E{-1}$ & $1.36\E{-2}$ & $1.34\E{-2}$ \\
16\,B & $1.73\E{2}$ & $1.05\E{1}$ & $1.06\E{1}$ & 1.0 & 1.0 & 1.0 \\
17\,A & $1.67\E{2}$ & $9.31\E{0}$ & $7.44\E{0}$ & $1.33\E{-1}$ & $2.88\E{-2}$ & $2.88\E{-2}$ \\
17\,B & $1.67\E{2}$ & $1.02\E{1}$ & $1.04\E{1}$ & 1.0 & 1.0 & 1.0 
\end{tabular}
\label{tab:magkappa}
\end{table*}
%
%
\end{document}